\documentclass[showpacs,superscriptaddress,10pt]{revtex4}
\usepackage{amssymb,amsmath}
\usepackage[dvips,final]{graphicx}
\usepackage{slashed}
\usepackage{url}
\usepackage{subfigure}
\usepackage{textcomp}
\usepackage{graphicx}
\usepackage{bm}
\usepackage[usenames]{color}

\usepackage{epsfig}

\input epsf.tex
\usepackage{amssymb}
\usepackage{amsmath}
\usepackage{bm}
\usepackage{amsthm}
\usepackage{braket}
\usepackage[title]{appendix}
\usepackage{amsopn}
\def\comment#1{}

\def\beq{\begin{equation}}
\def\eeq{\end{equation}}
\def\bea{\begin{eqnarray}}
\def\eea{\end{eqnarray}}
\def\d{\delta}
\makeatletter
\usepackage{ulem}
\usepackage{cancel}
\usepackage{color}

\usepackage{slashed}
\usepackage{bbm}
\usepackage[dvips,final]{graphicx}
\usepackage{url}
\usepackage{subfigure}
\usepackage{textcomp}
\usepackage{bm}
\usepackage{dcolumn}
\usepackage[usenames]{color}
\usepackage{ulem}
\usepackage{cancel}
\usepackage{amssymb}
\usepackage{amsmath}
\usepackage{bm}
\usepackage{amsthm}
\usepackage{amsopn}
\usepackage{epsfig}
\usepackage{amssymb}
\usepackage{amsmath}
\usepackage{bm}
\usepackage{amsthm}
\usepackage{braket}
\usepackage[title]{appendix}
\usepackage{amsopn}
\usepackage{hyperref}
\def\comment#1{}

\def\l{\left}
\def\r{\right}
\def\beq{\begin{equation}}
\def\eeq{\end{equation}}
\def\bea{\begin{eqnarray}}
\def\eea{\end{eqnarray}}

\def\d{\delta}

\def\p{\textbf{p}}
\def\q{\textbf{q}}
\def\Q{\textbf{Q}}
\def\wQ{\widetilde{\textbf{Q}}}
\def\k{\textbf{k}}

\def\tmic{t_{\textrm{mic}}}
\def\x{\textbf{x}}

\input epsf.tex
\def\d{\delta}

\def\comment#1{}


\begin{document}

\title{ Quantum Boltzmann equation for fermions: An attempt to calculate the NMR relaxation and decoherence times using quantum field theory techniques  }

\author{Hassan Manshouri}
\email[]{manshourihasan@gmail.com}

\affiliation{Department of Physics, Isfahan University of
Technology, Isfahan 84156-83111, Iran}

\author{Ahmad Hoseinpour}
\email[]{ahmad.hoseinpour@ph.iut.ac.ir}

\affiliation{Department of Physics, Isfahan University of Technology, Isfahan
84156-83111, Iran}
\affiliation{ICRANet-Isfahan, Isfahan University of Technology, 84156-83111, Iran}

\author{Moslem Zarei}
\email[]{m.zarei@iut.ac.ir}

\affiliation{Department of Physics, Isfahan University of Technology, Isfahan
84156-83111, Iran}
\affiliation{ICRANet-Isfahan, Isfahan University of Technology, 84156-83111, Iran}

\affiliation{Dipartimento di Fisica e Astronomia \textquotedblleft G. Galilei\textquotedblright, Universit\`{a} degli Studi di Padova, via Marzolo
8, I-35131, Padova, Italy}

\affiliation{INFN Sezione di Padova, via Marzolo 8, I-35131, Padova, Italy}

\date{\today}

\date{\today}

\begin{abstract}
Extracting macroscopic properties of a system from microscopic interactions has always been an interesting topic with the most diverse applications. Here, we use the quantum Boltzmann equation to investigate the density matrix evolution of a system of nucleons. Using the quantum field theory tools for constructing the density matrix operators and calculating the interactions is the main advantage of this equation. The right-hand side of this equation involves forward scattering and usual collision terms. As examples of application, we calculate the standard Bloch equations for the nucleon system in the presence of a constant and an oscillating magnetic field from the forward scattering term. We find the longitudinal and transverse (decoherence) relaxation times from the collision term by considering the nucleon-nucleon scattering.

\end{abstract}

%\pacs{13.15.+g, 98.80.Es, 98.70.Vc.}

\maketitle

 \section{Introduction}

Nuclear magnetic resonance (NMR) is an important tool with many applications in
different domains ranging from materials science to fundamental physics. NMR is a spectroscopy branch that can be used to measure the separation between two energy levels of a system. The accuracy of the measurement will increase considerably when the resonance of an applied radio-frequency (rf) field with the Larmor precession of the magnetic moments around a constant magnetic field is fulfilled. The motion of a system magnetic moment in a uniform magnetic field can be described either by a set of classical or quantum mechanical equations \cite{Berestetsky:1982aq}. According to the quantum theory, one can describe an ensemble of free spins by introducing a density matrix operator $\hat{\rho}$.
For a two-level system, two energy eigenstates $\l| +\r>$ (spin-up) and $\l| - \r>$ (spin-down), the density operator becomes a $2 \times 2$ density matrix $\rho$ with 3 independent Stokes parameters through which the Bloch vector is constructed \cite{Berestetsky:1982aq}. The master equation typically gives the time evolution of $\rho$. By writing the interaction Hamiltonian of the nucleon system with a magnetic field in the matrix form, substituting it into the master equation, one finds the system of torque equations that describe the Bloch vector evolution. However, the ideal torque equation changes due to the interaction of the spin system with the environment. The environmental noise induces two kinds of changes to the Bloch vector known in the literature as longitudinal relaxation ($T_1$ process) and transverse relaxation ($T_2$ process).
The longitudinal relaxation is concerned with the change of the diagonal element of the density matrix or equivalently the longitudinal component of the Bloch vector. The transverse relaxation is concerned with the decay of the off-diagonal (coherence) elements of the density matrix. Including both processes, the time evolution of the Bloch vector is given by a set of phenomenological equations called Bloch equations \cite{Abragam:1961}.

Furthermore, the dynamics of quantum two-level systems (the qubits) has always been of great interest, but recently attracted increasing attention due to its central importance for qubit-based quantum computation and quantum information processing in solid-state systems \cite{PhysRevLett.93.016601,2005PhRvA..71f0308T,PhysRevB.72.125337,PhysRevA.75.032333,PhysRevB.75.115307,PhysRevB.76.075333,Wilhelm_2008,2009JPCM...21M5602B}. The spin phase-coherence is of
central importance for spin-based quantum computation.
Sufficiently long coherence times with the ability to address the error correction schemes are needed for implementing quantum algorithms.
However, decoherence due to the coupling of a qubit to its environment is widely regarded as the major obstacle to the implementation of large-scale quantum computing.
In general, the environment-induced decoherence results from the unwanted interaction of a quantum system with its surrounding environment which, as a consequence, creates entanglement between the system and the environment. When such decoherence occurs, similar to the NMR systems, it is reflected in the temporal decay of off-diagonal elements of the reduced density matrix describing the system. The decoherence is crucial in describing the quantum-to-classical transition \cite{Schlosshauer:1105882}. However, this loss of coherence often significantly reduces the efficiency of quantum information protocols while a crucial requirement for quantum computers is the preservation of phase
coherence of a physical qubit subjected to a noisy environment. The environment is usually modeled by a bath of harmonic oscillators coupled to the qubits. However, to the decoherence time, it is crucial to find a complete microscopic description of the noise source as well as the interaction between the system and the environment.
Several many-body theories have been developed during recent years to describe the relaxation processes quantitatively and calculate the spin relaxation time $T_1$ and the spin decoherence time $T_2$ \cite{Levitt:2008,doi:10.1063/1.1840467,doi:10.1063/1.1669591,Yang_2016}.
 In this work, we introduce a new method to calculate the relaxation and decoherence times using the quantum Boltzmann equation (QBE). This equation is a powerful tool to study the effects of the microscopic interactions of the system with the environment on its mesoscopic time evolution. The QBE for photons has been developed by \cite{Kosowsky:1994cy,Alexander:2008fp, PhysRevD.81.084035, PhysRevD.98.023518, Bartolo:2019eac, PhysRevD.102.063501} in the context of the cosmic microwave background. The QBE has also been studied in great detail for a system of fermions in \cite{Fidler:2017pkg}. For this latter case, it is crucial to understand well the polarization of a fermion gas and its underlying classical description. However, there have been very few studies about the impacts of the forward scattering on fermion systems. To our knowledge, the only known example is flavor oscillations of neutrinos in the matter. It is shown that forward scattering is responsible for refractive effects such as the MSW effect \cite{Sigl:1992fn,Raffelt:1992uj}.
In QBE, we use quantum field theory (QFT) techniques to define the density operator and calculate the microscopic interactions. Quantum field theory is an essential tool for the quantum mechanical description
of processes that involve many-particles. The polarization matrix $\rho$ will be a central object in QBE. It is given by taking the expectation value over the number operator $\hat{\mathcal{D}}$ that is defined by multiplying a creation and an annihilation operator.
The number operator counts the number of particles in a given state in the Fock space. In QBE, the density operator is defined using the number operator $\hat{\mathcal{D}}$, so it basically contains all information about the system.
Another advantage of using QBE is the following. On the right-hand side of this equation, there are the forward scattering and collision terms written in terms of the commutators of the effective interaction Hamiltonian and the number operator.
The first term on the right-hand side of QBE describes forward scattering, and the second term describes nonforward interactions.
From these terms, we will be able to calculate the microscopic interaction of the system with the environment using the standard QFT techniques. Accordingly, after taking the expectation value, QBE relates the evolution of the macroscopic quantities that are in the density matrix to the microscopic interactions.
The forward and nonforward scattering terms are active during mesoscopic and microscopic time-scales.
The dynamics of such a classical system is evaluated during the mesoscopic time-scale.
In this paper, we calculate the relaxation and decoherence times using the techniques of QBE. The calculation method will be as follows. We consider a system of fermionic particles (nucleons) and assume that they interact through electromagnetic force. We also assume that an external magnetic field has been applied to this system. Under these conditions, the nucleon-nucleon scattering amplitude is calculated.  After taking the nonrelativistic limit, we decompose the scattering amplitude into the spin-spin and spin-orbit parts. Inserting these terms into the collision term of the QBE, we will calculate a system of coupled equations describing the time evolution of the Bloch vectors. By comparison these equations with the phenomenological equations obtained by Bloch \cite{blo:46}, we read the longitudinal and transverse relaxation times. Few papers have calculated the relaxation times using a scattering approach \cite{doi:10.1063/1.1840467,doi:10.1063/1.1669591,Yang_2016}.
The advantage of our approach is that one can explicitly determine the microscopic origin of relaxation and decoherence times. The following sections will discuss the relation between spin-spin and spin-orbit interactions with the longitudinal and transverse relaxation times. Our results can be applied to the systems containing monatomic gas with slow relaxation.

The paper is organized as follows:
In Sec. II, we will review and develop the basic structure of QBE for fermions. Using this equation, In Sec. III, we introduce a new interpretation of the NMR effect based on the forward scattering of nuclei from a background magnetic field. In Sec. IV, we calculate the relaxation times from the collision term of the QBE, and in Sec. V their magnitudes are estimated by considering a typical system of nucleons.  We summarize and conclude in Sec. VI.
We have also added two appendices, A and B, to extend the application of QBE to study the fermion-axion and the fermion-neutrino forward scattering.

%%%%%%%%%%%%%%%%%%%%%%%%%%%%%%%%%%

\section{Quantum Boltzmann Equation}\label{Boltzmann}

We consider a spin-$1/2$ system weakly coupled to an environment involving a finite number of spin-$1/2$ degrees
of freedom.
The Markovian QBE governs the evolution of the density matrix for such a system of fermions (f) out of equilibrium. Here, we first set up a density matrix corresponding to such a system of fermions. We consider an ensemble of spin-1/2 particles called system $\mathcal{S}$. The macroscopic properties of the system emerge by taking the expectation value of the density matrix. This quantity has a central role in QBE and is constructed as follows. We first write the Fourier transform of the quantum spinor field as
\beq \label{fourierpsi}
\psi^+(x)=\int \frac{d^3p}{(2\pi)^3}\sum_r\left[b_r(p)u_r(p)e^{-ip\cdot x}\right]~,
\eeq
and
\beq \label{fourierbarpsi}
\bar{\psi}^-(x)=\int \frac{d^3p}{(2\pi)^3}\sum_r\left[b^{\dag}_r(p)\bar{u}_r(p)e^{ip\cdot x}\right]~,
\eeq
where $m_f$ is the fermion mass, $u_r$ is the free spinor solution of the Dirac equation with spin index $r =1,2$ and $b_r(p)$ ($b_r^\dag(p)$) is the fermion annihilation (creation) operator obeying
the canonical anticommutation relation
\beq
\l\{b_r(p),b^\dag_{r'}(p')\r\}=(2\pi)^3\delta^3(\p-\p')\delta_{rr'}~.
\eeq
The density operator that describes the system of fermions is given by
\beq
\hat{\rho}^{(f)}=\int \frac{d^3 p'}{(2\pi)^3}\rho^{(f)}_{ij}(\p')b_i^\dag(\p')b_j(\p')~,
\eeq
where $\rho^{(f)}_{ij}$ is the fermion polarization matrix that describes the macroscopic properties of the system. The expectation value of a typical operator
$\hat{\mathcal{A}}$ can be extracted from $\hat{\rho}^{(f)}$ as
\beq
\l<\hat{\mathcal{A}}(\k)\r>=\textrm{tr}\l[\hat{\rho}^{(f)}\hat{\mathcal{A}}(\k)\r]=\int \frac{d^3\q'}{(2\pi)^3}\l<\q'  \left |\hat{\rho}^{(f)}\hat{\mathcal{A}}(\k) \right |\q'\r>~.
\eeq
 In particular, the expectation value of fermion number operator $\hat{\mathcal{D}}_{ij}^{(f)}(\k)=b_i^\dag(\k)b_j(\k)$ is given by
 \beq
\l<\hat{\mathcal{D}}^{(f)}_{ij}(\k)\r>=(2\pi)^3\delta^3(0)\rho^{(f)}_{ij}(\k)~,
\eeq
where $k^0$ is the energy of fermions. The polarization matrix $\rho^{(f)}$ of a system of nonrelativistic fermions can be parameterized using four Stokes parameters or Bloch vectors, and the representation of the SU(2) group \cite{Berestetsky:1982aq,Tolhoek:1956cy,Fidler:2017pkg}. We write
\begin{align}\label{density2}
\rho^{(f)}= \frac{1}{2}\left(\mathbb{I}+\bm{\sigma}\cdot\bm{\zeta}^{(f)}\right)=\frac{1}{2}\left(\begin{array}{cc}1+\zeta^{(f)}_z & \zeta^{(f)}_x-i\zeta^{(f)}_y \\ \zeta^{(f)}_x+i\zeta^{(f)}_y  & 1-\zeta^{(f)}_z\end{array}\right)~,
\end{align}
where $\mathbb{I}$ is the identity matrix, $\sigma_i$ are the Pauli spin matrices, $\zeta^{(f)}_z$,  $\zeta^{(f)}_x$ and $\zeta^{(f)}_y$ are the components of the Bloch vector.
The diagonal elements of the polarization matrix are called the populations. In particular,
$\zeta^{(f)}_z$ shows the difference in spin state populations, and indicates net longitudinal spin polarization or the system magnetization along the direction of the external magnetic field.

The number operator $\hat{\mathcal{D}}^{(f)}_{ij}(\mathbf{k})$ evolves according to the following equation \cite{Kosowsky:1994cy,Fidler:2017pkg}
 \begin{align}\label{boltzmann0}
\frac{d}{dt}\hat{\mathcal{D}}^{(f)}_{ij}(\mathbf{k},t)=i\l[H^{(n)}_{\textrm{int}}(t),\hat{\mathcal{D}}^{(f)}_{ij}(\mathbf{k},t)\r]-\frac{1}{2}\int_{-\infty}^{\infty} d\tmic\left[H^{(n)}_{\textrm{int}}(t),\left[H^{(n)}_{\textrm{int}}(t+\tmic),\hat{\mathcal{D}}^{(f)}_{ij}(\mathbf{k},t)\right]\right]+\mathcal{O}(g^{3n})~,
\end{align}
where $\tmic$ is the microscopic timescale quantifying the timescale of individual particle interactions.

Here, $H^{(n)}_{\textrm{int}}(t)$ is an effective interaction Hamiltonian which describes a scattering process of the fermion fields from a set of background fields.
 The superscript $n$ denotes the number of vertices in the corresponding Feynman diagrams of such process.
Each vertex corresponds to the fundamental interaction Hamiltonian $\mathcal{H}_{I}(g)$ in which $g$ is a general dimensionless coupling constant.
In general, $H^{(n)}_{\textrm{int}}(t)$ is defined in terms of the so called scattering operator $\hat{S}$ also known as the S-matrix. The scattering operator should be expressed in terms of the time-ordered exponential of $\mathcal{H}_{I}(t)$. We can write \cite{Kosowsky:1994cy}
\beq
\hat{S}=\sum_{n=0}^{\infty}\hat{S}^{(n)} \equiv \sum_{n=0}^{\infty} \frac{(-i)^n}{n!}\int d^4x_1 \cdot\cdot\cdot d^4x_n
T\left \{ \mathcal{H}_{I}(x_1) \cdot\cdot\cdot \mathcal{H}_{I}(x_n)\right \}~,
\eeq
where $T$ indicates the time-ordering.
 In this expression, $\hat{S}^{(n)}$ represents all scattering processes with $n$ vertices.
The $n$th term in the above series represents all scattering processes with $n$
interaction vertices. Now the effective interaction Hamiltonian $H^{(n)}_{\textrm{int}}(t)\sim g^{n}$ is defined as in the following
\beq
\hat{S}^{(n)} =\frac{(-i)^n}{n!}\int d^4x_1 \cdot\cdot\cdot d^4x_n
T\left \{ \mathcal{H}_{I}(x_1) \cdot\cdot\cdot \mathcal{H}_{I}(x_n)\right \} \equiv
-i \int_{-\infty}^{\infty} dt_1 H^{(n)}_{\textrm{int}}(t_1)~.
\eeq
Therefore, the scattering of fermion from an external background field for which $n=1$ is described by
\beq \label{s1}
\hat{S}^{(1)}=-i\int d^4x \,
 \mathcal{H}_{I}(x)
\equiv-i\int_{-\infty}^{\infty} dt \,H_{\textrm{int}}^{(1)}(t)~,
\eeq
and in the same way the fermion-fermion scattering of fermion for which $n=2$ is described by
\beq \label{s2}
\hat{S}^{(2)}= -\frac{1}{2}\int d^4x_1  d^4x_2\,
T\left \{ \mathcal{H}_{I}(x_1)  \mathcal{H}_{I}(x_2)\right \}
\equiv-i\int_{-\infty}^{\infty}  dt\, H_{\textrm{int}}^{(2)}(t)~.
\eeq

It is worth noting that \eqref{boltzmann0} is true up to $g^{2n}$ order in perturbation theory so the higher order terms have been dropped.
It is also important to note that the above equation has been obtained using the Markov approximation in which the timescale of the environment is taken to be much shorter than the timescale of the system so that the memory effects of the environment are negligible in the long run. We now take the expectation value of both sides of Eq. \eqref{boltzmann0} and arrive in the following equation called quantum Boltzmann equation for fermions \cite{Kosowsky:1994cy,Fidler:2017pkg}
 \bea
 \label{boltzmann1}
(2\pi)^3\delta^3(0) \frac{d}{dt}\rho^{(f)}_{ij}(\k,t) &=& i\l<\l[H^{(n)}_{\textrm{int}}(t),\hat{\mathcal{D}}^{(f)}_{ij}(\mathbf{k},t)\r]\r>-\frac{1}{2}\int_{-\infty}^{\infty} d\tmic \l<\left[H^{(n)}_{\textrm{int}}(t),\left[H^{(n)}_{\textrm{int}}(t+\tmic),\hat{\mathcal{D}}^{(f)}_{ij}(\mathbf{k},t)\right]\right]\r>
\nonumber \\ &+&
\mathcal{O}(g^{3n})~.
\eea
The first term in the right-hand side of the quantum Boltzmann equation \eqref{boltzmann1} is known as the forward scattering term, and the second term is known as the collision term \cite{Raffelt:1992bs,Fidler:2017pkg,Kosowsky:1994cy}. We will show that the forward scattering term is responsible for the NMR effect. From the collision term, we will extract the longitudinal and transverse (decoherence) relaxation times.
 In the following, we first study the froward scattering term and show that the NMR effect is due to the forward scattering of the nucleon with a background electromagnetic pulse. This effect causes the Bloch vectors to depart from thermal equilibrium. However, over a sufficiently long time, the thermal equilibrium state will be gradually reestablished due to the relaxation.
We will calculate the relaxation times from the electromagnetic interaction of the nucleon with nucleon surroundings.
%%%%%%%%%%%%%%%%%%%%%%%%%%%%%%%

\section{  Basic NMR effect from forward scattering of nucleon in an external electromagnetic field }

We consider a system of the spin-$1/2$ nucleons on which an external static magnetic field and a time-dependent, circularly polarized magnetic field are applied. The NMR phenomena rely on the interaction of the nuclear spin with the radio-frequency field. In general, the time evolution of this system's magnetization is given by a system of equations called Bloch equations after Felix Bloch \cite{blo:46}. Here, we will discuss that the Bloch equations can be extracted using the QBE formalism and the notion of the forward scattering of radiation from nucleons (N). Here, we will ignore the collision term to avoid the confusion.
\begin{figure}
  \centerline{\includegraphics[width=4cm]{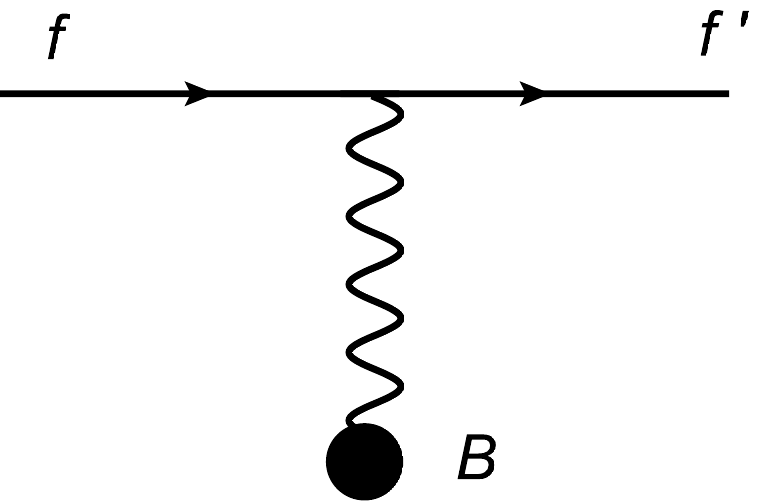}}
  \caption{The Feynman diagram for the fermion-external magnetic field forward scattering.
    }\label{Feyn-B}
\end{figure}

The nucleon interaction with an external electromagnetic field is shown in Fig. \ref{Feyn-B}. This scattering process is described by the following effective interaction Hamiltonian
\beq \label{Hint0}
\mathcal{H}_{I}= q_N\, \bar{\psi}(x)\gamma^{\mu}\psi(x)
\tilde{A}_{\mu}(x)~,
\eeq
where $q_N$ is the electric charge, $\psi(x)$ is spinor of nucleon and $\tilde{A}_{\mu}$ is the background electromagnetic field. Using the Gordon identity \cite{Peskin:257493}, one can decompose the interaction Hamiltonian as
\beq
\label{Hint1}
\mathcal{H}_{I}= \frac{q_N}{2M_N}\,\left[\left[i\bar{\psi}(x)\partial^{\mu}\psi(x)-i\partial^{\mu}\bar{\psi}(x)\psi(x)\right]\tilde{A}_{\mu}(x)+\frac{1}{2}\bar{\psi}(x)\sigma^{\mu\nu}\psi(x) \bar{F}^{\mu \nu}(x)\right]~,
\eeq
where $M_N$ is the neucleon mass and $\bar{F}^{\mu \nu}$ is the background electromagnetic strength tensor.
The second term in Eq. \eqref{Hint1} describes the interaction of a nucleon with an external magnetic field. This term is spin-dependent and in a nonrelativistic limit, gives the interaction of the external magnetic field with the magnetic moment of the nucleon. For the process shown in Fig. \eqref{Feyn-B}, using \eqref{s1} the interaction Hamiltonian is read as
\beq \label{hamiltonian:nnB0}
H^{(1)}_{\gamma_B N}(t)=\frac{q_N}{4M_N}\int d^3x\,\bar{\psi}^{-}(x)\sigma^{\mu\nu}\psi^{+}(x) \bar{F}_{\mu\nu}(x)~,
\eeq
where $\bar{\psi}^{-}$ and $\psi^{+}$ are linear in creation and absorption operators of nucleons respectively as shown in Eqs. \eqref{fourierpsi} and \eqref{fourierbarpsi}. It is also assumed that $\bar{F}_{\mu\nu}(x)$ is spatially uniform.
We now plug $H_{\textrm{int}}$ into the forward scattering term of the Eq. \eqref{boltzmann1} and using the Fourier transforms \eqref{fourierpsi} and \eqref{fourierbarpsi} we find
\bea
i\l<\l[H^{(1)}_{\gamma_B N}(t),\hat{\mathcal{D}}^{(f)}_{ij}(\mathbf{k},t)\r]\r>&=&\frac{iq_N}{4M_N}\sum_{rr'}\int d^3x\, d\p d\p'
 \bar{u}_{r'}(\mathbf{p}')\sigma^{\mu \nu}u_r(\mathbf{p})\bar{F}_{\mu\nu}(t)e^{i(\p-\p')\cdot\x} \nonumber \\ &\times&
 \l<b^{\dag}_{r'}(p')b_r(p)b^{\dag}_{i}(k)b_j(k)-b^{\dag}_{i}(k)b_j(k)b^{\dag}_{r'}(p')b_r(p)\r>~,
\eea
where
\beq
d\p=\frac{d^3p}{(2\pi)^3}~.
\eeq
The expectation values needed here are calculated using the Wick's theorem as below \cite{Kosowsky:1994cy}
\beq\label{expnucl2}
 \l<b^{\dag}_{r'}(p')b_r(p)b^{\dag}_{i}(k)b_j(k)\r>\simeq (2\pi)^6
 \delta^3(\p-\k)\delta^3(\k-\p')\,\delta_{ri}\,\rho^{(N)}_{jr'}(\k)~,
\eeq
On using the above expectation value and taking the integration over $x$, $\p$ and $\p'$ one simplifies the forward scattering term and plugging into the Boltzmann equation \eqref{boltzmann1} it follows that
\bea \label{boltzmannforward}
(2\pi)^3\delta^3(0) \frac{d}{dt}\rho^{(N)}_{ij}(\k,t)=i(2\pi)^3\delta^3(0)\sum_{rr'}\mathcal{M}(r,r')
 \l[\delta_{ri}\,\rho^{(N)}_{jr'}(\k,t)-\delta_{jr'}\,\rho^{(N)}_{ri}(\k,t)\r]+\mathcal{O}(g^{2})~,
\eea
where $g\sim q_N$ and $\mathcal{M}(r,r' )$ is the scattering matrix element of the process that is given by
 \begin{equation}\label{100}
  \mathcal{M}(r,r' ) = \frac{q_N}{4M_N}\bar{u}_{r'}(\mathbf{p'})\sigma^{\mu \nu}\bar{F}_{\mu\nu}(t)u_r(\mathbf{p})~,
\end{equation}
where $F^{\mu \nu}$ is the Fourier transform of the electromagnetic strength tensor. As it was mentioned above, here, we have ignored higher order terms of $\mathcal{O}(g^{2})$ to avoid the confusion.
 The spinor
$u_r(p)$ is the solution of the free Dirac equation \cite{Peskin:257493}
\beq  \label{urp}
u_r(p) =\left(
               \begin{array}{cc}
                 \sqrt{p\cdot\sigma}\,\chi_r \\
                  \sqrt{p\cdot\bar{\sigma}}\,\chi_r \\
               \end{array}
             \right)~,
\eeq
where $\bar{\sigma}^\mu=(1,-\vec{\sigma})$ and $\chi_r$ is the two-component spinor. We take the nonrelativistic limit and find the scattering amplitude in the form
\beq
\mathcal{M}(r,r') = - \chi^T_{r'}\,\bm{\mu}_N\cdot\mathbf{B}_{\textrm{tot}}(t)\,\chi_r~,
\eeq
where $T$ in the upper index means transposition, and $(B_{\textrm{tot}})_i(t)=-\epsilon_{ijk}\tilde{F}^{jk}(t)/2$ is the external magnetic field. In this expression, $\bm{\mu}_N$ denotes the nucleon magnetic moment
\beq
\bm{\mu}_N=g_N\frac{q_N}{2M_N} \mathbf{S}~,
\eeq
where $\mathbf{S}$ is the spin operator and $g_N$ is the nucleon gyromagnetic ratio.
The nuclear spin is placed in an external constant magnetic field $\mathbf{B}_0=B_0 \hat{z}$ and a periodic radio-frequency field $\mathbf{B}_{\textrm{rf}}(t)$ rotating in the clockwise direction and oriented as follows
\beq
\mathbf{B}_{\textrm{rf}}(t)=B_{\textrm{rf}}( \hat{x}\cos \omega t - \hat{y}\sin \omega t )~.
\eeq
 Most NMR experiments require that the magnetic field $\mathbf{B}_0$ be extremely homogenous and stable with respect to the time. Employing a superconducting magnet can provide such a homogeneous and stable magnetic field. The experiment also needs a radio-frequency field $\mathbf{B}_{\textrm{rf}}(t)$ with a very well-defined frequency. The total magnetic field $\mathbf{B}_{\textrm{tot}}(t)$ is
\beq
\mathbf{B}_{\textrm{tot}}(t)=B_{\textrm{rf}} \,\hat{x}\cos \omega t - B_{\textrm{rf}}\,\hat{y}\sin \omega t +B_0 \hat{z}~,
\eeq
 and therefore the scattering amplitude becomes
 \beq
 \mathcal{M}(r,r') = - \chi^T_{r'}\,\bm{\mu}_N\cdot \left(B_{\textrm{rf}} \,\hat{x}\cos \omega t - B_{\textrm{rf}}\,\hat{y}\sin \omega t +B_0 \hat{z}\right) \,\chi_r~.
 \eeq
Therefore, using this expression in the forward scattering term of Eq. \eqref{boltzmannforward} yields
\begin{equation}\label{102}
   \dot{\rho}^{(N)}_{ij}(\k,t)=-\frac{i}{2}\sum_{r'r}\chi^T_{r'}\bm{\sigma} \cdot\left( \hat{x}\,\omega_R\cos \omega t -\hat{y}\,\omega_R\sin \omega t + \hat{z}\,\omega_L\right)\chi_r \left(\delta_{ir}\rho^{(N)}_{r'j}(\mathbf{k},t)-\delta_{r'j}\rho^{(N)}_{ir}(\mathbf{k},t)\right)+\mathcal{O}(g^{2})~,
\end{equation}
in which we have defined the Larmor and the Rabi frequencies as follow \cite{bellac2006quantum}
\beq
\omega_L=g_N\frac{q_N}{2M_N}B_0~, ~~~~~~~~~~~  \omega_R=g_N\frac{q_N}{2M_N}B_{\textrm{rf}}~.
\eeq

From \eqref{density2} and \eqref{102} the evolution equations for $\zeta^{(N)}_i$ in the laboratory frame are given by
\bea
 \dot{\zeta}^{(N)}_x&=&\omega_R  \sin\omega t~ \zeta^{(N)}_z - \omega_L \zeta^{(N)}_y~,\label{zetax0}\\
\dot{\zeta}^{(N)}_y&=&-\omega_R \cos\omega t~  \zeta^{(N)}_z+\omega_L \zeta^{(N)}_x~,\label{zetay0}\\
  \dot{\zeta}^{(N)}_z&=&-\omega_R  \sin\omega t~ \zeta^{(N)}_x + \omega_R  \cos\omega t~ \zeta^{(N)}_y~.\label{zetaz0}
\eea

\begin{figure}
 \centerline{\includegraphics[width=4.5in,height=2in]{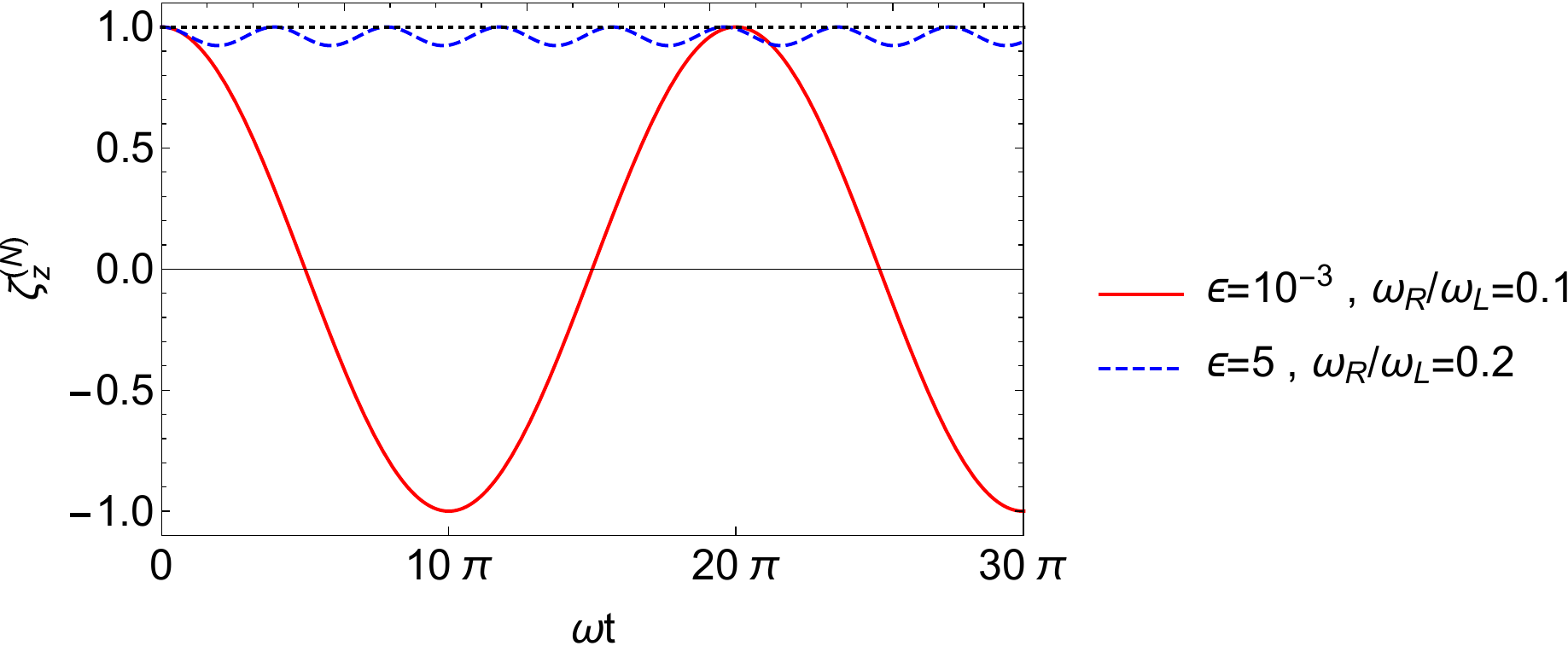}}
  \caption{The oscillation of the z component of the Bloch vector for a radiofrequency field in laboratory frame in the resonance condition where the difference between Larmor frequency and radiofrequency ($\omega$) is negligible in comparison with Rabi frequency ($\epsilon=(\omega-\omega_L)/ \omega_R\ll 1$) and off-resonance condition ($\epsilon=(\omega-\omega_L)/ \omega_R> 1)$.
    }\label{fig2}
\end{figure}

To find an analytical solution for $ \zeta^{(N)}_x $, we first eliminate the time-dependent of the magnetic field by defining two new Stokes parameters $\zeta^{(N)}_+$ and $\zeta^{(N)}_-$ in the following form
\beq
\left(\begin{array}{cc} \zeta^{(N)}_+\\ \zeta^{(N)}_- \end{array}\right)=
\left(\begin{array}{cc}\cos(\omega t) & \sin(\omega t) \\ -\sin(\omega t) & \cos(\omega t)\end{array}\right)
\left(\begin{array}{cc} \zeta_x\\ \zeta_y \end{array}\right)~.
\eeq
This is equivalent to use a coordinate system that rotates about the $z$-direction at frequency $\omega$. In such a coordinate system, $\mathbf{ B}_{\textrm{tot}}$ will be static.
Using the new basis, the system of differential equations is simplified as
\bea\label{stokes-fermion-manetic}
 \dot{\zeta}^{(N)}_-&=&- \Delta \omega\, \zeta^{(N)}_+ -\omega_R \zeta^{(N)}_z
 ~,\label{zetax0-1}\\
 \dot{\zeta}^{(N)}_+&=&  \Delta \omega\, \zeta^{(N)}_-~,\label{zetay0-2}\\
  \dot{\zeta}^{(N)}_z&=&\omega_R\, \zeta^{(N)}_-~,\label{zetaz0-3}
\eea
where $\Delta\omega=\omega -\omega _L$.
In the presence of the constant and time-dependent magnetic field, the equation of motion associated with the Bloch vector $\tilde{\bm{\zeta}}^{(N)}=( \zeta^{(N)}_+, \zeta^{(N)}_-, \zeta^{(N)}_z)$ can also be represented by the following classical equation \cite{Abragam:1961}
\beq \label{torque}
\dot{\tilde{\bm{\zeta}}}^{(N)}=\mu_N\,\tilde{\bm{\zeta}}^{(N)}\times  \mathbf{ B}_{\textrm{eff}}~,
\eeq
where
\beq
\mathbf{ B}_{\textrm{eff}}=\frac{1}{\mu_N}(-\omega_R,0,\Delta \omega)~,
\eeq
is an effective static field and $\mu_N=g_Nq_N/2M_N$ is the magnetic moment of nucleon.
In forward scattering, no energy and momentum are exchanged between the system and the environment. It turns out that the forward nucleon scattering with an applied magnetic field only causes the rotation of the nucleon polarization. Actually, as \eqref{torque} shows, forward scattering acts like a torque upon the nuclei. $\dot{\tilde{\bm{\zeta}}}^{(N)}$ is driving by torque and $\tilde{\bm{\zeta}}^{(N)}$ therefore precesses in a cone around the direction of $ \mathbf{ B}_{\textrm{eff}}$.

Now, taking the time derivative of Eq. \eqref{zetax0-1} and using the Eqs. \eqref{zetay0-2} and \eqref{zetaz0-3} we find a second order ordinary differential equations for $\zeta^{(N)}_-$ decoupled from  $\zeta^{(N)}_z$ and $\zeta^{(N)}_+$ as
\bea\label{second-zeta-n}
\frac{d^2 \zeta^{(N)}_-}{d t^2}=-\omega _R^2\left[1+\epsilon^2\right]\,\zeta^{(N)}_-~,
\eea
where $\epsilon=\Delta\omega/\omega_R$.
Solving this differential equation gives the following oscillatory solution
\beq
\zeta^{(N)}_-(t)=\frac{1}{\sqrt{1+\epsilon^2}}\,\zeta^{(N)}_z(0)\sin\left(\omega_R \sqrt{1+\epsilon^2} \,t\right)~,
\eeq
where $\zeta^{(N)}_z(0)$ is determined before imposing the external field $B_{\textrm{rf}}$.
Then we plug this solution into \eqref{zetay0-2} and \eqref{zetaz0-3}, so we find $\zeta^{(N)}_z$ and $\zeta^{(N)}_+$ in the following form
\beq \label{zetaz}
\zeta^{(N)}_z(t)=\frac{1}{1+\epsilon^2}\,\zeta^{(N)}_z(0)\left[\cos\left(\omega_R \sqrt{1+\epsilon^2} \,t\right)+\epsilon^2\right]~,
\eeq
and
\beq \label{zetap}
\zeta^{(N)}_+(t)=\frac{\epsilon}{1+\epsilon^2}\,\zeta^{(N)}_z(0)\left[\cos\left(\omega_R \sqrt{1+\epsilon^2} \,t\right)-1\right]~.
\eeq
In the condition that $\epsilon << 1$, the $\zeta^{(N)}_z$ oscillates with the Rabi frequency $\omega_R$ as
\beq
\zeta^{(N)}_z(t)=\zeta^{(N)}_z(0)\cos(\omega_R t)~.
\eeq
The red curve in Fig. \eqref{fig2} shows $\zeta^{(N)}_z(t)$ for $\epsilon=10^{-3}$ and $\omega_R/\omega_L=0.1$. In this case, one can find the maximum oscillation amplitude. This is the condition that is known as resonance.
Deviation from this limit reduces both oscillation amplitude and period.
The blue-dashed curve in Fig. \eqref{fig2} shows the $\zeta^{(N)}_z(t)$ for the case that $\epsilon=5$ and $\omega_R/\omega_L=0.2$.
This analysis is fully compliant with Rabi oscillation in NMR \cite{bellac2006quantum}.

The NMR signal is feeble and contains different sources of noise contributions.
So far, we have investigated the forward scattering of nucleons with a background of electromagnetic radiation.  The resulting Eqs. \eqref{zetax0}-\eqref{zetaz0} are a set of macroscopic equations that describe the Bloch vector or nuclear magnetization as a function of time.

It would also be quite interesting to explore some extensions and modifications of Eqs. \eqref{zetax0}-\eqref{zetaz0} by considering the forward scattering of nucleons with other backgrounds such as axionlike particles (ALPs) and neutrinos.
ALPs are one of the most attractive candidates for dark matter \cite{Abbott:1982af,Dine:1982ah,Preskill:1982cy,Wilczek:1987mv,Sikivie:1983ip,Sikivie:1985yu,Raffelt:1987im}. A variety of experiments designed to search ALPs based on weakly coupling of axions to fermions \cite{Irastorza_2018}. A large field occupation number of axion of the low momentum modes can act as a classical background field oscillating at a frequency $\omega_a$ equal to its mass $m_a$ \cite{Abbott:1982af,Dine:1982ah,Preskill:1982cy,Wilczek:1987mv,Sikivie:1983ip,Sikivie:1985yu,Raffelt:1987im,Irastorza_2018}. The coupling of the classical ALP field to nuclear or electronic spin gives rise to a precession of the spin that can change the magnetization of a sample of the material with a large number of spins. In principle, using the properties of fermions with a very high spin number like ferromagnetic crystals to search for new physics has been attracting increasing attention in recent years. In such experiments, it is expected that the interaction of fermions with the new physics sector will cause a coherent precession of the fermion spins. CASPEr-Wind experiment \cite{JacksonKimball:2017elr} searches the coupling of ALPs to the spin of the nucleus using NMR techniques. The nonrelativistic limit of the ALP-fermion interaction yields a Hamiltonian describing the spin operator coupling with the gradient of this time-varying background.
The QUAX experiment \cite{Barbieri:2016vwg} searches ALPs through their resonant coupling with the spin of a magnetized sample. As well as, a high precision SQUID magnetometer can observe the ALP-induced magnetization changes in this experiment. The detection of ALPs through their interactions with fermions may require the development of new experimental techniques. There may be many experimental methods that will be proposed in the near future. New schemes could potentially allow the detection of the QCD axion. Moreover, the improved techniques may provide the capability to probe hitherto unconstrained ALPs parameter space when the ALPs couples to fermions. All of these motivate us to develop a deeper theoretical understanding of the interaction of a beam or sample of fermions with other kinds of new particles such as ALP and neutrinos.
Accordingly, in Appendices A and B, we consider two further examples of the forward scattering of nucleons with axionlike particles and neutrinos, respectively. In Appendix A, we will derive the equation of Bloch vectors motion for a system of nucleons that interact with a time-dependent background of the axion field.
In Appendix B, we will consider the interaction of nucleons with neutrinos.

 \begin{figure}
  \centerline{\includegraphics[width=4cm]{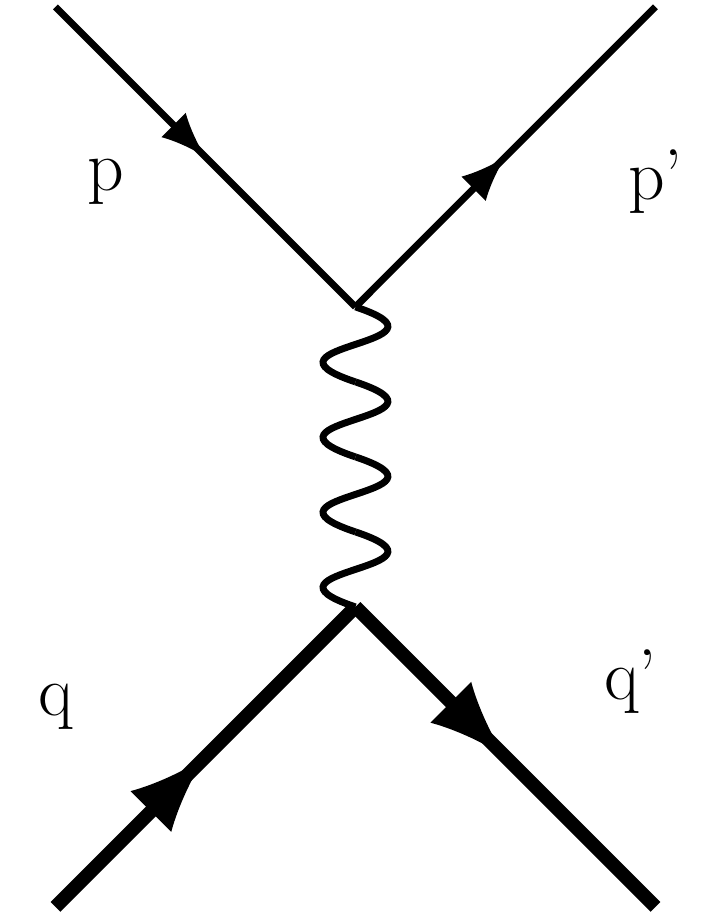}}
  \caption{The Feynman diagram for nucleon-fermion scattering.
    }
  \label{fig:3}
\end{figure}

 %%%%%%%%%%%%%%%%%%%%%%%%%%%

 \section{Calculating the scattering term: deriving relaxation and decoherence times}

In this section, we turn to the scattering term on the right-hand side of QBE. It should be noted that the precession of $\zeta^{(N)}_z(t)$ does not go on forever. There is a mechanism known as relaxation that relaxes the longitudinal and transverse components of the Bloch vectors. It is usually convenient to parameterize this mechanism using two phenomenological parameters known as longitudinal relaxation time $T_1$ and transverse decoherence time $T_2$ \cite{Levitt:2008}. They are responsible for modeling the drifting of the longitudinal
and the transverse component of the magnetization toward their thermal equilibrium values.
It is worth emphasizing that our results apply to an ensemble of free spin-$1/2$ fermions. In the following, we calculate $T_1$ and $T_2$ and then estimate them for free fermion gas.
Our observation shows the lifetime of a coherent superposition of spin up and spin down states must be sufficiently long.
To study the situation for the spin-$1/2$ fermions in solid-state systems one must assume that fermions are confined and expand the spinor fields in terms of new basses.
Here, we will not discuss such confined qubits any further but we will leave this for future investigation.

The extension of the Bloch equations by relaxation terms are given in the following form \cite{Abragam:1961}
\bea \label{b1}
 \dot{M}^{(N)}_x=\mu_N \l [\mathbf{ M}^{(N)}\times \mathbf{B}\r]_x-\frac{1}{T_2}M^{(N)}_x\\ \label{b11}
 \dot{M}^{(N)}_y=\mu_N \l[\mathbf{ M}^{(N)}\times \mathbf{B}\r]_y-\frac{1}{T_2}M^{(N)}_y\\ \label{b22}
 \dot{M}^{(N)}_z=\mu_N \l[\mathbf{ M}^{(N)}\times \mathbf{B}\r]_z-\frac{1}{T_1}\l[M^{(N)}_z-M^{(N)}_{\textrm{eq}}\r] \label{b33}
\eea
where $M^{(N)}_{\textrm{eq}}$ is the magnetization in the equilibrium condition and the Stokes parameters $\zeta^{(N)}_i$ is related to the magnetization $M^{(N)}_i$ of a system involving $N$ fermions in the following form
\bea \label{Mi}
M^{(N)}_i=N g_{N}\frac{q_{N}}{2 m_{N}} \zeta^{(N)}_i~.
\eea

In the large relaxation time limit, $T_1=T_2=\infty$, one recovers the standard equation \eqref{torque} induced by the forward scattering term.
  Here, we assume that the relaxation mechanism is due to the nucleon scattering from its surrounded particles, which are massive nucleons. Therefore, the time evolution of the Bloch vector will explicitly include new terms due to the interaction with the environment. Accordingly, we write a new effective interaction Hamiltonian describing this scattering and substitute it into the second term on the right-hand side of the Boltzmann equation \eqref{boltzmann1}.
We compare these new terms with the phenomenological Eqs. \eqref{b1}-\eqref{b22} and read the $T_1$ and $T_2$.
The scattering of a free nucleon from another free nucleon is described by the single Feynman diagram shown in Fig. \eqref{fig:3} (note that the nucleon-nucleon forward scattering vanishes). For this process, we write the corresponding S-matrix element and then find the effective interaction Hamiltonian using Eq. \eqref{s2}. The following Hamiltonian effectively describes the interaction
\beq \label{spin-spin0}
H^{(2)}_{\textrm{fN}}(t)= q_f q_N\int  d^3x d^4x'\, \bar{\psi}^{-}(x)
\gamma^{\mu}\psi^{+}(x)P_{\mu\nu}(x-x') \bar{\phi}^{-}(x')\gamma^{\nu}\phi^{+}(x')~,
\eeq
where we have assumed that both nucleons have the same electric charge and $P_{\mu\nu}(x-x')$ is the conventional Feynman propagator for the photons
\beq
P_{\mu\nu}(x)=-i\int \frac{d^4k}{(2\pi)^4}\frac{g_{\mu\nu}}{k^2+i\epsilon}\,e^{-ik\cdot x}~,
\eeq
and $\phi(x)$ is the spinor field of the surrounding fermions (nuclei) that is given in terms of in terms of absorbtion and creation operators as
\beq \label{fourierbarphi}
\phi^{+}(x)=\int \frac{d^3q}{(2\pi)^3}\sum_s d_s(q)U_s(q)e^{-iq\cdot x}~,
\eeq
\beq \label{fourierphi}
\bar{\phi}^{-}(x)=\int \frac{d^3q}{(2\pi)^3}\sum_s  d^{\dag}_s(q)\bar{U}_s(q)e^{iq\cdot x}~,
\eeq
where $U_s(q)$ is a spinor defined as \eqref{urp} for a fermion with mass $m_f$, and $d^{\dag}$ and $d$ are the fermions creation and annihilation operators satisfying the following canonical  anticommutation relation
\beq
\left \{d_s(p),d^{\dag}_{s'}(q')\right \}=(2\pi)^3\delta^3(\q-\q')\d_{ss'}~,
\eeq
and the expectation value of creation and annihilation operators are given as
\bea
\braket{d^\dagger_{s'}(q')d_s(q)}=(2\pi)^3 \delta^3(\mathbf{q}-\mathbf{q'}) \rho^{(f)}_{s' s}(\q)~,
\eea
and
\bea\label{expferm2}
\braket{d^\dagger_{s'_1}(q'_1)d_{s_1}(q_1)d^\dagger_{s'_2}(q'_2)d_{s_2}(q_2)}&=&(2\pi)^6 \delta^3(\mathbf{q_1}-\mathbf{q'_1})\delta^3(\mathbf{q_2}-\mathbf{q'_2}) \rho^{(f)}_{s'_1 s_1} \rho^{(f)}_{s'_2 s_2}
\nonumber \\& & \:\:
+(2\pi)^6 \delta^3(\mathbf{q_1}-\mathbf{q'_2})\delta^3(\mathbf{q_2}-\mathbf{q'_1}) \rho^{(f)}_{s'_1 s_2}(\q_2) \left[\delta_{s_1 s'_2}-\rho^{(f)}_{s'_2 s_1}(\q_1)\right]~,
\eea
where $\rho^{(f)}_{ij}$ is the polarization matrix of the surroundings (environment).  In the presence of an external magnetic field, the spin of fermions interact with the magnetic field. In this case, the spin is fully aligned in the direction of the magnetic field. However, when $T\neq 0$, thermal fluctuations impede the system from fully aligning with the magnetic field. The degree of alignment can be quantified by the
magnetization $\rho^{(f)}_{11}(\q)-\rho^{(f)}_{22}(\q)$. The number density of fermions in the bulk is also given by $\rho^{(f)}_{11}(\q)+\rho^{(f)}_{22}(\q)=n^{(f)}(\q)$. In the same way, the expectation values of the operators associated with the nucleon system is given by
\bea\label{expneucl2}
\braket{b^\dagger_{r'_1}(p'_1)b_{r_1}(p_1)b^\dagger_{r'_2}(p'_2)b_{r_2}(p_2)}&=&(2\pi)^6 \delta^3(\mathbf{p_1}-\mathbf{p'_1})\delta^3(\mathbf{p_2}-\mathbf{p'_2}) \rho_{r'_1 r_1}(\p_1) \rho_{r'_2 r_2}(\p_2)
\nonumber \\& & \:\:
+(2\pi)^6 \delta^3(\mathbf{p_1}-\mathbf{p'_2})\delta^3(\mathbf{p_2}-\mathbf{p'_1}) \rho_{r'_1 r_2}(\p_2) \left[\delta_{r_1 r'_2}-\rho_{r'_2 r_1 }(\p_1)\right]~.
\eea
The interaction Hamiltonian \eqref{spin-spin0}, in principle, involves spin-orbit and spin-spin interaction terms.
To decompose these interactions, we employ the Gordon identity in \eqref{spin-spin0} and write
\bea \label{spin-spin1}
H^{(2)}_{\textrm{fN}}(t)&=&
\frac{q_{N} q_f}{4m_f M_N}\int d^3x d^4x' \left[i\bar{\psi}^-(x)\partial^{\mu}\psi^+(x)-i\partial^{\mu}\bar{\psi}^-(x)\psi^+(x)+\frac{1}{2}\partial_{\alpha}\left(\bar{\psi}^-(x)\sigma^{\mu\alpha}\psi^+(x)\right)\right] \nonumber \\& \times &
P_{\mu\nu}(x-x')\left[i\bar{\phi}^-(x')\partial^{\nu}\phi^+(x')-i\partial^{\nu}\bar{\phi}^-(x')\phi^+_s(x')+\frac{1}{2}\partial_{\beta}\left(\bar{\phi}^-(x')\sigma^{\nu\beta}\phi^+(x')\right)\right]
~.
\eea
The Fourier transform of Hamiltonian is obtained by substituting \eqref{fourierpsi}, \eqref{fourierbarpsi} , \eqref{fourierphi} and \eqref{fourierbarphi},  into \eqref{spin-spin1}. Therefore, after taking integrating over $x$ and $x'$ the resulting interaction Hamiltonian in Fourier space becomes
\bea \label{spin-spin2}
H^{(2)}_{\textrm{fN}}(t)&=&
\int   d\p d\p'd\q d\q' \sum_{r,r',s,s'}e^{-i(p^0+q^0-p'^0-q'^0)t}
\mathcal{M}(pr,p'r',qs,q's')\nonumber\\ &\times &
(2\pi)^3\delta^3(\p+\q-\p'-\q')b^{\dag}_{r'}(p')b_{r}(p)d^{\dag}_{s'}(q')d_{s}(q)
~,
\eea
where $\mathcal{M}$ expresses the scattering amplitude
\bea\label{MGordon}
\mathcal{M}(q's',p'r',qs,pr)&=&\frac{q_{N}q_f}{4m_fM_N}\left[(p+p')^\mu \bar{u}_{r'}(p')u_r(p)+\frac{i}{2}(p'-p)_\alpha\bar{u}_{r'}(p)\sigma^{\mu\alpha}u_r(p)\right] \nonumber \\& \times&
 \frac{g_{\mu\nu}}{(p-p')^2}\left[(q+q')^\nu \bar{U}_{s'}(q')U_s(q)+\frac{i}{2}(q'-q)_\beta\bar{U}_{s'}(q)\sigma^{\nu\beta}U_s(q)\right]~,
\eea
and we use the following abbreviations
\beq
d\p= \frac{d^3 p}{(2\pi)^3}~,~~~~~~~ d\q= \frac{d^3 q}{(2\pi)^3}~.
\eeq
We are interested in the nonrelativistic limit in which the spinors are expanded as in the following
\bea
u_r(\p) \simeq \left(\begin{array}{cc} \chi_r\\ \frac{\bm{\sigma}\cdot\p}{2M_N} \chi_r \end{array}\right)~,
\eea
where, $\chi_r$ and $\chi_{r'}$ are the two-component spinors and $\bm{\sigma}$ are the Pauli matrices. Applying this approximation, the expressions within the brackets of Eq. \eqref{MGordon} are simplified as
 \bea
 \bar{u}_{r'}(\p')\left[p'^0+p^0+i \sigma ^{0 \alpha } \left(p'_{\alpha }-p_{\alpha }\right)\right] u_{r}(\p)\simeq \chi _{r'}^*\left(M_N+\frac{\p'^2}{2 M_N}+\frac{\p^2}{2M_N}\right)\chi _{r}+\chi _{r'}^*\frac{i\bm{\sigma}\cdot \Q\times \p}{2 M_N}\chi _{r}~,
\eea
and
\bea
 \bar{u}_{r'}(\p')\left[p'^j+p^j+i \sigma ^{j \alpha} \left(p'_{\alpha }-p_{\alpha }\right)\right] u_{r}(\p) \simeq \chi _{r'}^*\left[2p'^j+ Q^j+i\left(\Q\times \mathbf{\bm{\sigma} }\right)^j\right]\chi _{r}~,
 \eea
where $\Q=\p'-\p=\q'-\q$ is the transfer momentum. Finally, after taking the low-energy limit
the matrix elements associated with the spin-orbit and spin-spin terms are given in the following form
\bea\label{spin-orbit}
& &\mathcal{M}_{\textrm{spin-orbit}}= \nonumber\\
& &q_f q_N\l\{\chi _{r'}^*\left[\frac{i\bm{\sigma}\cdot\Q\times \p}{4M_N^2|\Q|^2}-\frac{i\bm{\sigma} \cdot\Q\times \q}{2m_f M_N|\Q|^2}\right] \chi _{r}\chi _{s'}^* \chi_s
+ q_f q_N\chi _{r'}^* \chi _{r}\chi _{s'}^*\left[\frac{i\bm{\sigma} \cdot\Q \times \p}{2m_f M_N|\Q|^2}-\frac{i\bm{\sigma}\cdot\Q\times \q}{4m_f^2|\Q|^2}\right]\chi _s \r\}~,
\eea
and
\bea\label{spin-spin}
\mathcal{M}_{\textrm{spin-spin}}=\frac{q_f q_N}{4m_f M_N|\Q|^2}\,\left(\chi _{r'}^*i\bm{\sigma }\times \Q \chi _{r}\right) \cdot \left(  \chi _{s'}^*i\bm{\sigma}\times \Q \chi _s\right)~.
\eea
One can also check that our results for \eqref{spin-orbit} and \eqref{spin-spin} are consistent with \cite{Berestetsky:1982aq}. Accordingly, the interaction Hamiltonian \eqref{spin-spin2} consists of two parts
 \bea
 H^{(2)}_{\textrm{fN}}= H_{\textrm{spin-orbit}}+H_{\textrm{spin-spin}}~.
 \eea
Now, we substitute the interaction Hamiltonian into the scattering term of QBE \eqref{boltzmann1} and using \eqref{expneucl2} and \eqref{expferm2} we take the expectation value. In general, the collision term can be decomposed in the following nonzero two scattering terms
\bea
\mathcal{C}_{ij}(\k,t)&=& \mathcal{C}^{\textrm{spin-orbit}}_{ij}(\k,t)+ \mathcal{C}^{\textrm{spin-spin}}_{ij}(\k,t)~,
\eea
where
\beq \label{c1int}
 \mathcal{C}^{\textrm{spin-orbit}}_{ij}(\k,t)=-\frac{1}{2}\int_{-\infty}^{\infty} d\tmic \l<\left[H_{\textrm{spin-orbit}}(t),\left[H_{\textrm{spin-orbit}}(t+\tmic),\hat{\mathcal{D}}^{(N)}_{ij}(\mathbf{k},t)\right]\right]\r>~,
\eeq

\beq  \label{C-spin-spin}
 \mathcal{C}^{\textrm{spin-spin}}_{ij}(\k,t)=-\frac{1}{2}\int_{-\infty}^{\infty} d\tmic \l<\left[H_{\textrm{spin-spin}}(t),\left[H_{\textrm{spin-spin}}(t+\tmic),\hat{\mathcal{D}}^{(N)}_{ij}(\mathbf{k},t)\right]\right]\r>~.
\eeq
The cross terms that would involve one factor of $H_{\textrm{spin-orbit}}(t)$ and one factor of $H_{\textrm{spin-spin}}(t)$ vanish.
In the following, we evaluate all the scattering terms $\mathcal{C}^{\textrm{spin-orbit}}_{ij}$ and  $ \mathcal{C}^{\textrm{spin-spin}}_{ij}$ and study their impacts on the relaxation and dephasing times, individually.

%%%%%%%%%%%%%%%%%%%%%%%%

\subsection{Spin-orbit interaction}

Substituting $H_{\textrm{spin-orbit}}$ in \eqref{c1int}, using the expectation values \eqref{expferm2} and \eqref{expneucl2} and after taking the integration over $\p'$ and $\q'$ we find
\bea \label{C-spin-orbit}
& &\mathcal{C}^{\textrm{spin-orbit}}_{ij}(\k,t)=\nonumber\\
& &-\frac{\pi}{4}\int d\mathbf{q}d\mathbf{p}
\delta\left(E(-\k+\p+\q)+k^0-E\left(\q\right)-E\left(\p\right)\right)\sum_{\textrm{spins}}\mathcal{M}_{\textrm{spin-orbit}}(1)\mathcal{M}_{\textrm{spin-orbit}}(2)
\nonumber \\& & \:\:\:\:\:\:\:\:\:
\times
\left\{\delta _{s_1 s'_2}\rho^{(f)}_{s'_1 s_2} (\q )\delta _{r_1r'_2}\left[\delta _{i r_2}\rho^{(N)} _{r'_1j}\left(\k,t\right)+\delta _{jr'_1}\rho^{(N)} _{i r_2}\left(\k,t\right)\right]-2\delta _{s_2 s'_1}\rho^{(f)}_{s'_2 s_1}(-\k+\p+\q)\delta _{i r_2}\delta _{jr'_1}\rho^{(N)} _{r'_2r_1}(\p,t)\right\}~,\nonumber\\
\eea
where the argument of the matrix elements indicates the subscript to be attached to all dependent variables. Using \eqref{spin-orbit} and \eqref{C-spin-orbit} we get
\bea\label{M-spin-orbit}
\mathcal{M}_{\textrm{spin-orbit}}(1)\mathcal{M}_{\textrm{spin-orbit}}(2)&=&
q_f^2 q_N^2
\l\{
-\chi _{r'_1}^*\frac{i\bm{\sigma}\cdot\p\times \k}{4M_N^2|\wQ|^2} \chi _{r_1}\chi _{s'_1}^* \chi _{s_1}\chi _{r'_2}^*\frac{i\bm{\sigma}\cdot\p\times \k}{4M_N^2|\wQ|^2} \chi _{r_2} \chi^*_{s'_2}\chi _{s_2}
\r. \nonumber\\&+& \l.
\chi _{r'_1}^*\frac{i\bm{\sigma}\cdot\p\times \k}{4M_N^2|\wQ|^2} \chi _{r_1}\chi _{s'_1}^* \chi _{s_1}\chi _{r'_2}^*\frac{i\bm{\sigma}\cdot\wQ\times \q}{2m_f M_N \wQ^2} \chi _{r_2} \chi^*_{s'_2}\chi _{s_2}
\r. \nonumber\\&+& \l.
\chi _{r'_1}^*\frac{i\bm{\sigma}\cdot\wQ\times \q}{2m_f M_N|\wQ|^2} \chi _{r_1}\chi _{s'_1}^* \chi _{s_1}\chi _{r'_2}^*\frac{i\bm{\sigma}\cdot\p\times \k}{4M_N^2|\wQ|^2} \chi _{r_2} \chi^*_{s'_2}\chi _{s_2}
\r. \nonumber\\&+& \l.
\chi _{r'_1}^*\frac{i\bm{\sigma}\cdot\wQ\times \q}{2m_f M_N|\wQ|^2} \chi _{r_1}\chi _{s'_1}^* \chi _{s_1}\chi _{r'_2}^*\frac{i\bm{\sigma}\cdot\wQ\times \q}{2m_f M_N|\wQ|^2} \chi _{r_2} \chi^*_{s'_2}\chi _{s_2}\r \}~,
\eea
with $\wQ=\p-\k$. We now insert \eqref{M-spin-orbit} into the scattering term \eqref{C-spin-orbit} and take the integration over $\p$. Moreover, using this fact that in the nonrelativistic limit, that $|\q|\ll m_f$, $|\wQ|\ll m_f$ and $|\wQ|\ll |\q|$ we can further simplify the expressions in \eqref{C-spin-spin} by expanding the Dirac delta function and the energy $E$ in terms of $|\q|/m_f$ and $|\wQ|/m_f$ as in the following form
\bea \label{Ek}
E(-\k+\p+\q)&\approx& m_f \l(1+ \frac{|\q|^2}{2m^2_f}+ \frac{|\wQ|^2}{2m^2_f}+ \frac{\q\cdot\wQ}{m^2_f}+ \cdot\cdot\cdot\r)
\nonumber\\&\approx & E(\q)+ \frac{\q\cdot\wQ}{m_f}
~,
\eea
and
\bea  \label{delta}
\delta\left(E(-\k+\p+\q)+k^0-E\left(\q\right)-E\left(\p\right)\right) &\approx&
 \delta\left(E(\k)-E(\q)+ \frac{\q\cdot\wQ}{m_f}\right)
\nonumber\\ &\approx &
\delta(E(\k)-E(\p))-\frac{\q \cdot\wQ}{m_f}\frac{\partial}{\partial E(\p)}\delta(E(\k)-E(\p))
\nonumber\\ &\approx &
 \delta(E(\k)-E(\p))~,
\eea
where in the last step we have neglected the corrections of order $|\wQ|/ m_f$.
By substituting \eqref{M-spin-orbit} in \eqref{C-spin-orbit} and using \eqref{Ek} and \eqref{delta} we find the time evolution of density matrix
\bea\label{rhoij3}
\dot{\rho }^{(N)}_{i j}(\k,t)=&-&q_f^2 q_N^2\frac{\pi}{8  M_N^4} \int  d\q d\p \, \delta\left(E\left(\k\right)-E\left(\p\right)\right)\frac{1}{\wQ^4}
\l \{n^{(f)}\left(\q \right)\rho^{(N)} _{i j}\left(\k,t\right)\left|\p \times \k\right|^2
 \r. \nonumber\\&-& \l.
n^{(f)}\left(\q+\p-\k\right)\sum_{r r'} \rho^{(N)} _{r'r}(\p,t)\l[ \chi^*_j\bm{\sigma}\cdot \left(\p\times \k\right) \chi _{r}\r]
\l[\chi^* _{r'}\bm{\sigma}\cdot \left(\p\times \k\right) \chi _i \r] \r \} \nonumber\\
&-&q_f^2 q_N^2\frac{\pi}{2 m_f^2 M_N^2}\int d\q d\p \,\delta \left(E\left(\k\right)-E\left(\p\right)\right)\frac{1}{\wQ^4}
\l \{n^{(f)}\left(\q\right)\rho^{(N)} _{i j}\left(\k,t\right)\left|\wQ \times \q \right|^2
\r. \nonumber\\&-& \l.
n^{(f)}\left(\q+\p-\k\right)\sum_{r r'} \rho^{(N)} _{r'r}(\p,t)\l[\chi^* _j\bm{\sigma}\cdot \left(\wQ \times \q\right) \chi _{r}\r] \l[\chi _{r'}^*
\bm{\sigma}\cdot \left(\wQ \times \q\right) \chi _i \r ] \r \}~.
\eea

During the relaxation process, the rf magnetic field is switched off and the spins start realigning themselves back in a low energy state or equilibrium state that is determined through the main magnetic field's direction $\mathbf{ B}_0$.
In the presence of an external constant magnetic field $B_0$, it is expected that the spin system reaches the thermal equilibrium with its surroundings. In this situation, for the diagonal part of the polarization matrix, the ratio of $\rho^{(f)}_{11}$ and $\rho^{(f)}_{22}$ is given by the Boltzmann law \cite{Blum:1433745}
\beq \label{ratio}
\frac{\rho^{(f)}_{11}}{\rho^{(f)}_{22}}=\frac{\exp(-\beta E_1)}{\exp(-\beta E_2)},
\eeq
where $\beta=1 / T$ (we set Boltzmann constant $k_B = 1$), $E_1$ is the energy of the ground state $\l|1\r>$ with a spin aligned in the direction of $B_0$ , and $E_2$ is the energy of the excited state $\l|2\r>$  with a spin aligned in the opposite direction of $B_0$. Introducing the Boltzmann factor in \eqref{ratio} indicates that there is a larger population in state $\l|1\r>$.
The off-diagonal components are known as the coherence. They represent the transverse spin magnetization, i.e., a net spin polarization perpendicular to the external magnetic field.
Accordingly,  the density matrix elements for a system of Nuclei are weighted by transition rates $W_i$ \cite{Blum:1433745}.
For a two level system the transition rate from the state $\l|1\r>$ to the state $\l|2\r>$ is $W_1$ and from $\l|2\r>$ to $\l|1\r>$ is $W_2$. Therefore, elements of the density matrix is defined by
\bea
\rho^{(N)}= \left(\begin{array}{cc}W_1\rho_{11} & \frac{1}{2}W\rho_{12} \\ \frac{1}{2}W\rho_{21}  & W_2 \rho_{22}\end{array}\right)~,
\eea
where $W=W_1+W_2$ is the transition rate for the population that for the coherence density matrix elements is $W/2$~. In the presence of a static external magnetic field $B_0$, the transition rates are $W_1=\exp(-\mu_N B_0/2T)$ and $W_2=\exp(\mu_N B_0/2T)$ \cite{Yang_2016}.

The basis for the nucleon directions are also taken to be
\bea
\hat{\mathbf{ k}}&=&\l(\cos \phi_0 \sin \theta_0 ,\sin \phi_0 \sin \theta_0 ,\cos \theta_0 \r)~, \\
\hat{\mathbf{ p}}&=&\l( \cos \phi \sin \theta,\sin \phi \sin \theta ,\cos \theta \r) ~.
\eea
We now take the integrations over $\p$ and $\q$ and find the time evolution of $\zeta^{(N)}_i$ parameters as follows
\bea \label{zeta3-1}
\dot{\zeta }^{(N)}_x(\k,t)&=&-q_f^2 q_N^2
\l\{ \frac{1}{8 \pi  M_N}\l[\frac{ |\k| }{4  M^2_N}  \cos 2\phi _0\left(\cot ^2 \theta _0- \cos 2\phi _0\right)+\frac{1}{| \k|}\frac{ T}{m_f} \frac{ \cos 2\phi _0}{\sin^2\theta _0}\r]n^{(f)}(\x)\frac{W}{2} \zeta^{(N)}_x(\k,t)
\r. \nonumber\\&+& \l.
\frac{1}{32 \pi }\frac{ |\k| }{  M^3_N} \cot \theta _0  \cos 2\phi _0 \cos \phi _0 n^{(f)}(\x)W (\zeta^{(N)}_z (\k,t)-\zeta^{(N)}_{\textrm{eq}})
\r \}~,
\eea
\bea  \label{zeta3-2}
\dot{\zeta }^{(N)}_y(\k,t)&=&-q_f^2 q_N^2
\l \{ \frac{1}{8 \pi  M_N}\l[\frac{ |\k| }{4  M^2_N}  \cos 2\phi _0\left(\cot ^2 \theta _0- \cos 2\phi _0\right)+\frac{1}{| \k|}\frac{ T}{m_f} \sin\phi_0\frac{ \cos 2\phi _0}{\sin^2\theta _0}\r]n^{(f)}(\x)\frac{W}{2} \zeta^{(N)}_y(\k,t)
\r. \nonumber\\&+&\l.
\frac{1}{32 \pi }\frac{ |\k| }{  M^3_N} \cot \theta _0  \cos 2\phi _0 \sin \phi _0 n^{(f)}(\x)W (\zeta^{(N)}_z (\k,t)-\zeta^{(N)}_{\textrm{eq}})
\r \} ~,
\eea
\bea  \label{zeta3-3}
\dot{\zeta }^{(N)}_z(\k,t)&=&q_f^2 q_N^2\frac{| \k|}{32 \pi  M_N^3} n^{(f)}(\x) \frac{W}{2} \cos 2\phi _0 \cot \theta _0\left [-\cos \phi _0 \zeta^{(N)}_x\left(\k,t\right)+\sin \phi _0 \zeta^{(N)}_y\left(\k,t\right)\right ] ~,
\eea
where $\zeta^{(N)}_{\textrm{eq}} = (W_1-W_2)/W$ and $n^{(f)}(\x)$ is the number density of the surrounding fermions that is given by
\bea
\int \frac{d^3 q}{(2\pi)^3} n^{(f)} (\x,\q)=n^{(f)}(\x)~.
\eea
 Note that all the terms proportional to the inverse of $|\k|$ are generated from the second integration in Eq. \eqref{rhoij3}. The $\hat{\k}$-dependence of $\zeta^{(N)}_i$ is canceled by taking the integration of all directions $\hat{\k}$ in Eqs. \eqref{zeta3-1}-\eqref{zeta3-3}. In this case, all terms proportional to $|\k|$ are removed. We are interested in the time evolution equations of the magnetizations $M^{(N)}_i$~, so that using the relation \eqref{Mi} we can rewrite Eqs. \eqref{zeta3-1}-\eqref{zeta3-3} in terms of magnetization of the system
\bea\label{Mspin-orbitX}
\dot{M }^{(N)}_x=\frac{q_f^2 q_N^2}{16 \pi M^2_N}\sqrt{ \frac{T}{2\pi M_N}} \,n^{(f)}(\x)\frac{W}{2} M^{(N)}_x~,
\eea
\bea\label{Mspin-orbitY}
\dot{M }^{(N)}_y=\frac{q_f^2 q_N^2}{16 \pi M^2_N}\sqrt{ \frac{T}{2\pi M_N}} \,n^{(f)}(\x)\frac{W}{2} M^{(N)}_y~,
\eea
\bea\label{Mspin-orbitZ}
\dot{M }^{(N)}_z=0~,
\eea
where here we have ignored the forward scattering terms.
As we will see in the next sections, the spin-spin interaction makes a contribution which is larger in magnitude and opposite in sign compared to the spin-orbit interaction. Combining the effects of spin-spin and spin-orbit interactions lead to effective transverse relaxation. Moreover, the right-hand side of Eq. \eqref{Mspin-orbitZ} vanishes that implies that spin-orbit interaction does not generate longitudinal relaxation.

%%%%%%%%%%%%%%%%%%%%%%%%%%%%

\subsection{Spin-spin interaction}

Here, we focus on the spin-spin interaction of nucleon and surrounding fermions and calculate its corresponding collision term \eqref{C-spin-spin}.
Substituting $H_{\textrm{spin-spin}}$ in \eqref{C-spin-spin}, using the expectation values \eqref{expferm2} and \eqref{expneucl2} and taking the integration over $\p'$ and $\q'$ the spin-spin collision term becomes
\bea \label{spin-spin222}
& &\mathcal{C}^{\textrm{spin-spin}}_{ij}(\k,t)=\nonumber\\
& & -q_f^2 q_N^2\frac{1}{16}(2\pi)^4 \delta^3(0)\int d\mathbf{q}d\mathbf{p}
\delta\left(E(-\k+\p+\q)+k^0-E(\q)-E(\p)\right)\mathcal{M}_{\textrm{spin-spin}}(1)\mathcal{M}_{\textrm{spin-spin}}(2)
\nonumber \\
& & \times \left\{\delta _{s_1 s'_2}n_{s'_1 s_2} (\q)\delta _{r_1r'_2}\left[\delta _{i r_2}\rho^{(N)} _{r'_1j}\left(\k,t\right)+\delta _{jr'_1}\rho^{(N)} _{i r_2}\left(\k,t\right)\right]-2\delta _{s_2 s'_1}n_{s'_2 s_1}(-\k+\p+\q)\delta _{i r_2}\delta _{jr'_1}\rho^{(N)} _{r'_2r_1}(\p,t)\right\}~,
\eea
where the matrix elements are evaluated as in the following from
\bea\label{spin-spinspin-spin}
& & \mathcal{M}_{\textrm{spin-spin}} (1)\mathcal{M}_{\textrm{spin-spin}} (2)=\nonumber\\
& &\frac{q_f^2 q_N^2}{16^2 m_f^2 M_N^2|\wQ|^4}\l\{\l[\chi _{r'_1}^*\wQ\cdot\bm{\sigma } \chi _{r_1}\r]\l[ \chi^* _{s'_1}\wQ\cdot\bm{\sigma }\chi _{s_1}\r]  \l[  \chi _{r'_2}^*\wQ\cdot\bm{\sigma } \chi _{r_2}\r]\l[ \chi^* _{s'_2}\wQ\cdot\bm{\sigma }\chi _{s_2}\r]
 \r. \nonumber \\&-& \l.
   |\wQ|^2 \l [\chi _{r'_1}^*\wQ\cdot\bm{\sigma } \chi _{r_1} \r] \l[ \chi^* _{s'_1}\wQ\cdot\bm{\sigma }\chi _{s_1}\r]\l[ \chi _{r'_2}^*\sigma_m \chi _{r_2}\r]\l[ \chi _{s'_2}^*\sigma_m \chi _{s_2}\r] -
|\wQ|^2 \l[\chi _{r'_1}^*\sigma_m \chi _{r_1}\r]\l[ \chi _{s'_1}^*\sigma_m \chi _{s_1}\r]\l[\chi _{r'_2}^*\wQ\cdot\bm{\sigma } \chi _{r_2} \r]
\r. \nonumber \\&\times&  \l.
\l[ \chi^* _{s'_2}\wQ\cdot\bm{\sigma }\chi _{s_2}\r]+
|\wQ|^4 \l [\chi _{r'_1}^*\sigma_m \chi _{r_1} \r ] \l[ \chi _{s'_1}^*\sigma_m \chi _{s_1}\r] \l[\chi _{r'_2}^*\sigma_n \chi _{r_2} \r]\l[ \chi _{s'_2}^*\sigma_n \chi _{s_2}\r]
\r\}~,
\eea
In the following, we will investigate the contributions of the first term and the other three remaining terms of \eqref{spin-spinspin-spin} in the scattering terms separately. Substituting the above result in \eqref{spin-spin222} and taking the integration over $\q$ and $\p$ yields
\bea\label{rhospin-spin}
 \dot{\rho }^{(N)}_{i j}(\k,t)&=&
 -q_f^2 q_N^2\frac{\pi}{8 m_f^2 M_N^2}\int d\q d\p \delta \left(E(\k)-E(\p)\right)\l \{n^{(f)}(\q)\rho^{(N)} _{i j}(\k,t)
 \r. \nonumber \\&-&  \l.
 \frac{1}{|\wQ|^2}n^{(f)}(\q+\p-\k)\sum_{rr'} \l[\chi^* _j \wQ\cdot\bm{\sigma }\chi _{r}\r]\l[\chi^* _{r'}\wQ\cdot\bm{\sigma }\chi _i\r]\,\rho^{(N)} _{r'r}(\p,t)\r \}
 \nonumber\\
&+ &q_f^2 q_N^2\frac{\pi}{8 m_f^2 M_N^2} \int d\q d\p \delta \left(E\left(\k\right)-E(\p)\right)  \l \{2n^{(f)}(\q)\rho^{(N)} _{i j}(\k,t)
 \r. \nonumber \\&+&  \l.
n^{(f)}\left(\q+\p-\k\right)\sum_{rr'}\l[
\l[\chi^*_j{\bm{\sigma }}_n\chi _{r}\r] \l[\chi^* _{r'}{\bm{\sigma }}_n\chi _i\r]\rho^{(N)} _{r'r}(\p,t)
\r.\r. \nonumber \\&-&  \l.\l.
2\frac{1}{|\wQ|^2}\l[\chi^*_j \wQ\cdot\bm{\sigma }\chi _{r}\r]\l[\chi^* _{r'}\wQ\cdot\bm{\sigma }\chi _i\r] \,\rho^{(N)} _{r'r}(\p,t)
\r ]\r\}~,
\eea
in which the first integration includes the contribution of the first term of \eqref{spin-spinspin-spin}, and the second integration consists of the other three terms.
Now, We find the Bloch equations associated with the first integrations in Eq. \eqref{rhospin-spin}. The procedure is similar to the previous section.
We first obtain the evolution of Stokes parameters $\zeta^{(N)}_i$ and then find the Bloch equations by integrating over momentum $\k$. In the end, using the relation \eqref{Mi} we find
\bea\label{Mspin-spinX}
\dot{M}^{(N)}_x=-\frac{q_f^2 q_N^2}{4\pi m_f^2 } \sqrt{\frac{T}{2 \pi M_N}}\,n^{(f)}(\x)\frac{W}{2} M^{(N)}_x~,
\eea
\bea\label{Mspin-spinY}
 \dot{M }^{(N)}_y=-\frac{q_f^2 q_N^2}{4\pi m_f^2 } \sqrt{\frac{T}{2 \pi M_N}}\,n^{(f)}(\x)\frac{W}{2} M^{(N)}_y~,
\eea
\bea\label{Mspin-spinZ}
\dot{M }^{(N)}_z=-\frac{q_f^2 q_N^2}{4\pi m_f^2 } \sqrt{\frac{T}{2 \pi M_N}}\,n^{(f)}(\x)W \l[ M^{(N)}_z-M^{(N)}_{\textrm{eq}}\r]~,
\eea
where $M^{(N)}_{\textrm{eq}}=g_N\frac{q_N}{2 M_N}\zeta^{(N)}_{\textrm{eq}}$ is the magnetization in thermal equilibrium. For a magnetic field $B_0$ aligned in the $z$-direction and in the limit that $T \gg \frac{g_N q_N}{2 M_N}B_0$, we have
\bea
M^{(N)}_{\textrm{eq}}=N \l(\frac{g_Nq_N}{2 M_N}\r)^2 \frac{B_0}{ T} ~.
\eea
Therefore, the contribution of the first term of Eq. \eqref{spin-spinspin-spin}, leads to the transverse and the longitudinal relaxations. The contribution of the remaining three terms of matrix density in \eqref{spin-spinspin-spin} leads to the following equations
\bea\label{spin-spin21}
\dot{M}^{(N)}_x=-\frac{q_f^2 q_N^2}{2\pi m_f^2 } \sqrt{\frac{T}{2 \pi M_N}}\,n^{(f)}(\x)\frac{W}{2} M^{(N)}_x\nonumber\\
\eea
\bea\label{spin-spin22}
 \dot{M }^{(N)}_y=-\frac{q_f^2 q_N^2}{2\pi m_f^2 } \sqrt{\frac{T}{2 \pi M_N}}\,n^{(f)}(\x)\frac{W}{2} M^{(N)}_y\nonumber\\
\eea
\bea\label{spin-spin23}
\dot{M }^{(N)}_z=0~.
\eea
As a result, the contributions of theses terms do not generate the longitudinal relaxation. In the following section, we will derive the relaxation times due to spin-orbit and spin-spin interactions.

%%%%%%%%%%%%%%%%%%%%%%%%%%%%

\section{Discussion  }

 The main  feature of the Bloch equation is that the spin relaxation processes effect is described in terms of two real parameters: relaxation time $T_1$ and decoherence time $T_2$. As we discussed above, the spin relaxations of a system of nucleons can be extracted using QBE and the interaction of nucleons with surrounding fermions.
Decoherence due to the coupling of a qubit to its environment is widely regarded as one of the major obstacles to
quantum computing using solid-state systems. Here, we consider unconfined spin systems. We have shown that these systems couple to their environments primarily through the spin-orbit interaction and spin-spin interaction with nuclear spins in the surrounding lattice.
We separated the spin-spin and spin-orbit interactions to keep track of the physical nature of $T_1$ and $T_2$.
We have shown only the spin-spin interaction for the nucleon system could produce longitudinal relaxation. However, both interactions cause decoherence.

We start with the longitudinal relaxation time $T_1$ responsible for the relaxation of the diagonal elements of the polarization matrix. The new feature of our analysis is that all terms arising from the spin-orbit interaction eventually eliminate each other. The only nonvanishing contribution is due to the spin-spin interaction and the first term of \eqref{spin-spinspin-spin}.
Comparing Eq. \eqref{Mspin-spinZ}  with the phenomenological equation \eqref{b22}, we can read $T_1$ as in the following form
\bea\label{T1}
\frac{1}{T_1} =\frac{q_f^2 q_N^2}{2\pi m_f^2} \sqrt{\frac{ T}{2\pi M_N}}\, n^{(f)}(\x)~.
\eea
By definition, spin decoherence refers to the phenomena that tend to destroy the off-diagonal elements of the polarization matrix. To read the decoherence time $T_2$ we must first combine the Eqs. \eqref{Mspin-orbitX}-\eqref{Mspin-orbitZ} due to the spin-orbit interaction and also Eqs. \eqref{Mspin-spinX} - \eqref{spin-spin23} that are due to the spin-spin interactions. It is also conventional to introduce the pure decoherence time $T_{\varphi}$ that is due to those interaction terms that do not change the diagonal components of the polarization matrix.
Therefore, from Eqs. \eqref{Mspin-orbitX}-\eqref{Mspin-orbitZ} and \eqref{spin-spin21}-\eqref{spin-spin23} we find $T_{\varphi}$ as in the following
\beq
\frac{1}{T_{\varphi}} =\frac{ q_f^2 q_N^2}{16\pi } \l( \frac{8}{m_f^2}-\frac{1}{M_N^2}\r ) \sqrt{\frac{ T}{2\pi M_N}}\, n^{(f)}(\x)~.
\eeq
As a result, the total decoherence time is given by \cite{Yang_2016}
\beq
\frac{1}{T_2}=\frac{1}{2T_1}+\frac{1}{T_{\varphi}}~.
\eeq
Using these simple formulas the magnitude of $T_1$ and $T_2$ can be easily estimated. If the surrounding is nucleons that means $m_f=M_N\simeq 936\, \textrm{MeV}$, and for a typical sample like He$^3$ with the number density $n^{(f)} \simeq 10^{21}\, \textrm{cm}^{-3}$ and $T=4\, \textrm{K} $ \cite{doi:10.1063/1.1679671} we get
\beq \label{t1-1}
T_1\simeq10^4 \,\,\textrm{sec.}~,
\eeq
and
\beq \label{t2-1}
T_2 \simeq \frac{2}{3} T_1~.
\eeq
The resulting relaxation times indicate slow relaxation convenient in monatomic gases \cite{Abragam:1961,doi:10.1063/1.1679671}.
On the other hand, in the case that the interaction of electrons with mass $m_f\simeq 0.5\, \textrm{MeV}$ with the system of nucleus is noticeable, the resulting relaxation times become very different form \eqref{t1-1} and \eqref{t2-1}. For the Hydrogen gas in which electron-nucleon interaction is dominated we obtain
\beq \label{t1-2}
T_1\simeq10^{-1} \,\,\textrm{sec.}~,~~~\textrm{and}~~~ T_2 \simeq \frac{2}{3} T_1~,
\eeq
where $n^{(f)} \simeq 10^{19}\, \textrm{cm}^{-3}$ and $T=20\, \textrm{K} $ \cite{Abragam:1961,doi:10.1063/1.1679671}.
The relaxation times are linearly dependent on the number density of surroundings. For lower number densities, one finds a longer relaxation process.
In summary, it is only one term due to the spin-spin interaction that produces longitudinal relaxation. However, both spin-spin and spin-orbit interactions lead to transverse relaxation. The relaxation times that are calculated due to these interactions are in the same order.

 The QBE technique can also be applied to the case of solid-state two-level systems such as electron spins confined in
semiconductor quantum dots to investigate the decoherence due to the coupling of the system to its environment.
Careful modeling of the system-environment coupling in solid-state systems allows making theoretical predictions for the relaxation time $T_1$  and the decoherence time $T_2$. For such systems, the predicted relaxation $T_1$ time of spin is consistent with experiments. It is also known that the decoherence time $T_2$ can be much smaller than $T_1$,
although it's upper bound is $T_2 \leqslant  2T_1$. In particular, it would be very interesting to formulate self-consistent theories to predict the detailed origin of the $T_2$.

%%%%%%%%%%%%%%%%%%%%

\begin{acknowledgments}

We would like to thank Soroush Shakeri, Farhang Loran, Mehdi Abdi, Giovanni Carugno,  and Sabino Matarrese for their useful comments and suggestions. M.Z. acknowledges financial support by the University of Padova under the MSCA Seal of Excellence @UniPD programme.

\end{acknowledgments}

%%%%%%%%%%%%%%%%%%%%%%%%%%%%%%%%%%%%%%%%%%%%%%%%%%%%%%%%%%%%%%%%%%%%%%%%%%%%%%%%%%

\appendix

\section{Interaction with axionlike particles background}

In this Appendix, consider the forward scattering of the nucleon from the time-dependent axion field and use the QBE \eqref{boltzmann1} to re-derive the equation of motion of Bloch vector.
As it was discussed, a large field occupation number of ALPs quanta of the low momentum modes can exhibit a classical background field oscillating at a frequency $\omega_a$ that is equal to its mass $m_a$ \cite{Preskill:1982cy}. The coupling of the classical axion field to the nucleon spin gives rise to a precession of the spin \cite{Graham:2013gfa} that can change the magnetization of a sample with a large number of spins.

\begin{figure}
  \centerline{\includegraphics[width=4cm]{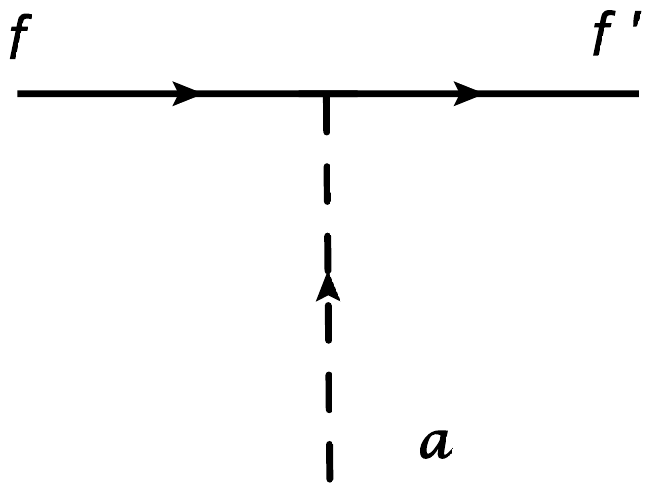}}
  \caption{The Feynman diagram for a fermion with an axion field.
    }
  \label{fig:5}
\end{figure}

Interactions of the axion field $a(x)$ generates a potential energy density $m^2_a a^2(x) /2 $, where $m_a$ is the axion mass. Oscillations of this field from the minimum of this potential cause the Compton frequency of axion field as $\omega_a= m_a$.
It is also known that the axion field behaves as a periodic classical background due to its large
occupation number. Therefore, it is assumed that the axion field has a gradient as
\beq
a(\mathbf{x},t)=a_0 A(\mathbf{x}) \cos(\omega_at)~,
\eeq
where $a_0$ is the amplitude of field with the dimension of energy, $A(\mathbf{x})$ is a dimensionless nonsingular position-dependent. Now, we assume a system of nucleons interacting with this oscillating background field.  Possible coupling of the nucleon with axion is in the following form \cite{Budker:2013hfa}
\beq
\mathcal{H}_{I}=g_{a}\partial_{\mu}a\bar{\psi}_N\gamma^{\mu}\gamma^{5}\psi_N~,
\eeq
where $g_a$ is the interaction coupling constant. Feynman diagram of the interaction for a fermion with axion is shown in Fig. (\ref{fig:5}). Using Eq. \eqref{s1} we can find the effective interaction Hamiltonian $H^{(1)}_{\textrm{int}}$ that describes this process. After plugging the Fourier transforms of the nucleon field in $H^{(1)}_{\textrm{int}}$ we have
\beq
H^{(1)}_{\textrm{Na}}=-g_{a}\int d^3 x \int d\mathbf{p}\int d\mathbf{p}' \partial_{\mu}a(x)\,\bar{u}_{r'}(p')\gamma^{\mu}\gamma^{5}u_r(p) b^{\dag}_{r'}(p')
b_r(p)e^{i(p'-p)\cdot x}~.
\eeq
Using the Gordon identity \cite{Peskin:257493} we have
\beq
\bar{u}_{r'}(p')\gamma^{\mu}\gamma^{5}u_r(p)=\frac{1}{2M_N}\,\bar{u}_{r'}(p')\left[(p'-p)^{\mu}\gamma^{5}+i\sigma^{\mu\nu}(p+p')_{\nu}\gamma^{5}\right]u_r(p)~,
\eeq
and assuming that nucleon is in the rest mass frame,
we get
\beq
H^{(1)}_{\textrm{Na}}=-g_{a}\int d^3 x \int d\mathbf{p}\int d\mathbf{p}' \partial_{i}a\,\bar{u}_{r'}(p')i\sigma^{i0}u_r(p) b^{\dag}_{r'}(p')
b_r(p)~.
\eeq
Inserting the above result into the forward scattering term of QBE \eqref{boltzmann1} and taking the integration over momenta, we find
\beq \label{rhodot1}
\dot{\rho}^{(N)}_{ij}(\k,t)=g_{a}\, a_0 \,\chi^{T}_{r'}\left<\bm{\nabla}A\right>\cdot \bm{\sigma}\,\chi_r\cos(\omega_a t)\left(\delta_{ri}\rho^{(N)}_{r'j}(\k,t)-\delta_{jr'}\rho^{(N)}_{ir}(\k,t)\right)~,
\eeq
where $\left<\bm{\nabla}A\right>$ denotes the average of the gradient term as
\beq
 \left<\bm{\nabla}A\right>=\frac{1}{V}\int d^3 x  \bm{\nabla}A~.
\eeq
This expression can be correspond to localized lumps of the axion field (commonly known as soliton) held together by their self-interaction. This comprehensive solution represents a stable configuration with a nonvanishing averaged gradient.
The motion of the Earth through the galaxy leads to a relative velocity known as axion ``wind" between it and this configuration.
  We can define an effective oscillating magnetic field induced by the axion field as
\beq
\mathbf{ G}_a=\frac{g_{a} }{\mu_N} a_0
 \left<\bm{\nabla}A\right>~.
\eeq
For a scalar field theory in the vacuum state one can generally write $a_0=\frac{\sqrt{2\rho_a}}{m_a}$ where $\rho_a$ is the energy density of axion.
Therefore, the Eq. \eqref{rhodot1} can be expressed as spin-magnetic field interaction described in the previous section. We also consider an external constant magnetic field $\mathbf{B}_0$ and hence
\beq \label{rhodot2}
\dot{\rho}^{(N)}_{ij}(\k,t)=\chi^{T}_{r'}\bm{\sigma}_N\,\chi_r\cdot\mu_N( \mathbf{G}_a\cos(\omega_a t)+\mathbf{ B}_0)\left(\delta_{ri}\rho^{(N)}_{r'j}(\k,t)-\delta_{jr'}\rho^{(N)}_{ir}(\k,t)\right)~.
\eeq
Using this equation one can find the time evolution of Bloch vectors. To this end and in order to use NMR techniques, we assume the initial condition $\bm{\zeta}^{(N)}(0)=(0,0,1)$~, and align the external magnetic field $\mathbf{ B}_0$ in this direction. The direction of $ \mathbf{ G}_a$ is in any direction that is not collinear with $\bm{\zeta}^{(N)}(0)$. Here, we assume that $ \mathbf{ G}_a=G_a \hat{\mathbf{x}}$.
With these conditions, the Boltzmann equation is given by
 \bea\label{eqaxion}
   \dot{\zeta}^{(N)}_x&=&-2 \omega_L \,\zeta^{(N)}_y~,\\
  \dot{\zeta}^{(N)}_y&=& 2\omega_L \zeta^{(N)}_x - 2\omega_R \cos(\omega_a t)\,\zeta^{(N)}_z~,\\
   \dot{\zeta}^{(N)}_z&=& 2\omega_R \cos(\omega_a t)\,\zeta^{(N)}_y~,
\eea
where $\omega_L=\mu_N B_0$ and $\omega_R=\mu_N G_a$. Much like the NMR effect described above, the current coupling will cause spin precession of a fermion around the local direction of $\mathbf{ G}_a$ as long as the fermion spin is not aligned with it.

%%%%%%%%%%%%%%%%%%%%%%%%%%

\section{Interaction with neutrino}

\begin{figure}
  \centerline{\includegraphics[width=4in,height=2in]{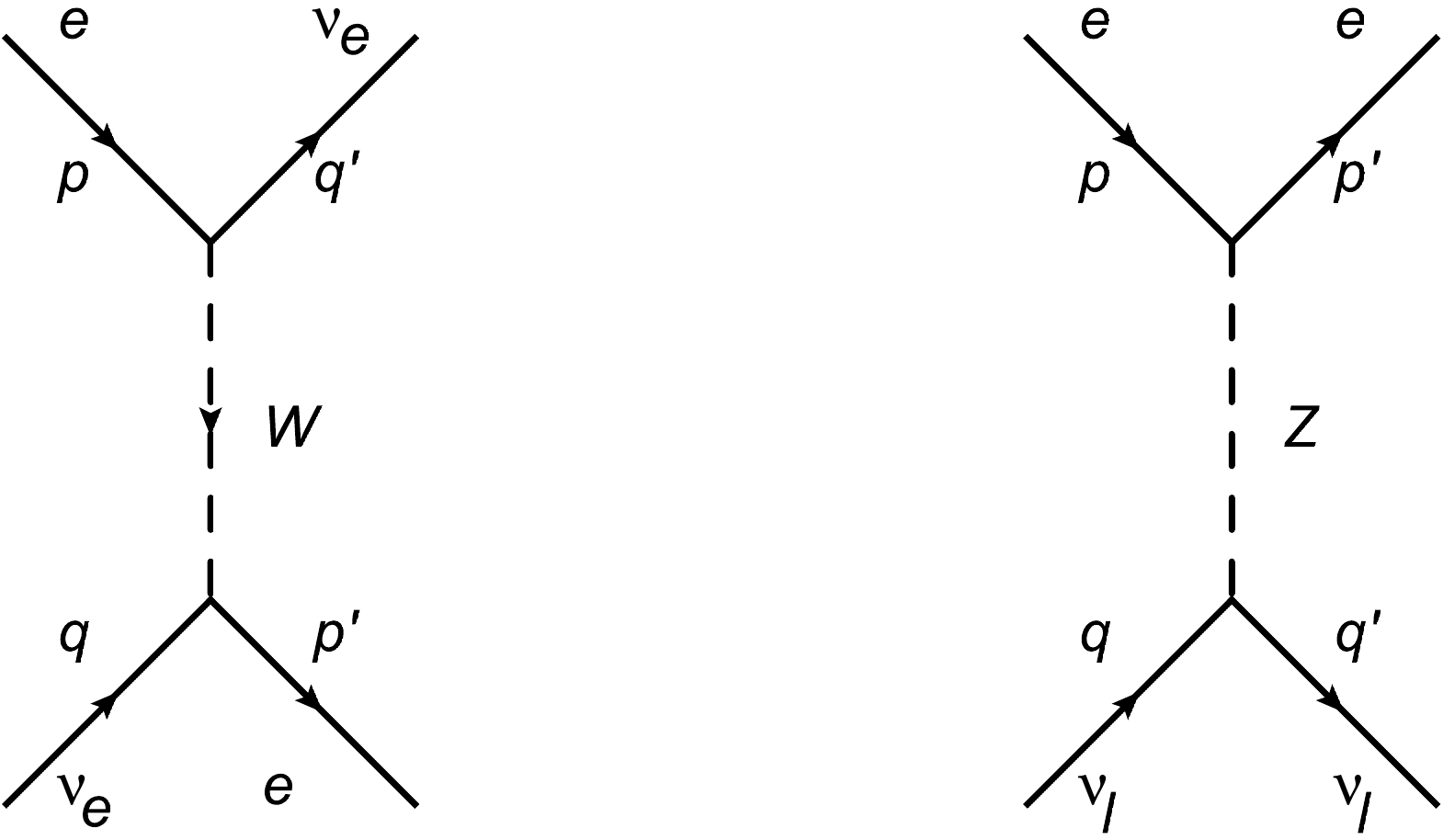}}
  \caption{The weak interaction Feynman diagrams for nucleon-neutrino forward scattering.
    }
  \label{fig:1}
\end{figure}

As another example, we calculate the effect of nucleon-neutrino forward scattering on the Bloch vector's evolution. Neutrino interacts weakly with nucleons in the framework of the standard model of particle physics. The two lowest-order Feynman diagrams for the process $n+ \nu \rightarrow n+ \nu$ are shown in Fig. \eqref{fig:1}. The left diagram is the $t-$channel charged current scattering process mediated by $ W $ particle exchange, and the right diagram shows the $t$-channel neutral current scattering process mediated by $ Z $ particle exchange. We can write the S-matrix element for this process and then find the interaction Hamiltonian using the Eq. \eqref{s2}. In general, the interaction Hamiltonian describing this process is given in the following form
\bea\label{H2in}
&&H^{(2)}_{\textrm{N}\nu}(t)=\int d\mathbf{q}d\mathbf{q}'d\mathbf{p}d\mathbf{p}'(2\pi)^3\delta^3(\mathbf{q}'+\mathbf{p}'-\mathbf{q}-\mathbf{p})\exp[it(q'^0+p'^0-q^0-p^0)]\nonumber \\&& \:\:\:\:\:\:\:\:\:\:\:\:\:\:\:\:\:\:\:\:\:\:\:\:\:\:\:\times
[d^{\dagger}_{r'}(q')b^{\dagger}_{s'}(p')\mathcal{M}(n(s,p)+\nu(r,q)\rightarrow n(s',p')+\nu(r',q'))\,b_s(p)d_r(q)]~,
\eea
 where $\mathcal{M}$ is the scattering matrix element and $d_r(q)$ and $d^\dagger_{r}(q)$ are the neutrino annihilation and creation operators, respectively. One can also write the following expectation value
\begin{align}\label{expd}
\braket{d^\dagger_{r'}(q') d_r(q)}=(2\pi)^3 q^0 \delta^3 (\mathbf{q}-\mathbf{q'}) \delta_{rr'} \frac{1}{2}\, n_\nu (\mathbf{x},\mathbf{q})~,
\end{align}
where $n_\nu (\mathbf{x},\mathbf{q})$ is the number density of neutrinos of momentum $\mathbf{q}$ per unit volume.
The matrix element associated with the $W$-exchange diagram is given by
\begin{align}\label{eq:20}
\mathcal{M}(ps,p's',qr,q'r')=-\frac{m^2_{\textrm{W}}G_F}{\sqrt{2}}\left[ \bar{u}_{s'}(p') \gamma ^ \mu(1-\gamma^5) u_r(q) \,\frac {g_{\mu\nu}+(q-p')_{\mu} (q'-p)_{\nu}/m^2_{\textrm{W}}}{(q'-p)^2-m^2_{\textrm{W}} } \,\bar{u}_{r'}(q') \gamma^\nu (1-\gamma^5) u_s(p)\right]~,
\end{align}
where $G_F$ is Fermi constant, $u_{s}$ are the nucleon spinors and $v_{r}$ are the neutrino spinors. Substituting this matrix element in the forward scattering term of the Eq. (\ref{boltzmann1}), we find
\begin{align}
\frac{d\rho^{(N)}_{ij}(\mathbf{k},t)}{dt}=-i\frac{G_F}{2\sqrt{2}}\int d\mathbf{q} \,n_{\nu}(\mathbf{x,q})  \left(\delta_{is} \rho^{(N)}_{s'j}(\k,t)- \delta_{ js'} \rho^{(N)}_{is}(\k,t) \right)
  \left[ \bar{u}_{s'}(k) \gamma _ \mu(1-\gamma^5) u_r(q) \bar{u}_{r}(q) \gamma^\nu (1-\gamma^5) u_s(k)\right]~,\label{Boltzmann3}
\end{align}
The neutrino spinor is defined according to its helicity in the ultrarelativistic limit \cite{Thomson:2013zua}. Using these helicity states, one can show by some straightforward calculations that the right-hand side of Eq. \eqref{Boltzmann3} vanishes. Therefore, the $W$-exchange diagram has no contribution to the evolution of the Bloch vector.

We now turn to the Z-exchange diagram. In the low energy limit, the scattering matrix element is given by
\begin{align}
\mathcal{M}(ps,p's',qr,q'r')=\frac{G_F}{4\sqrt{2}}\left[\bar {u}_{r'}(q')\gamma^\mu (1-\gamma
^5) u_r(q)\right]\left[\bar {u}_{s'}(p')\gamma_{\mu} ( c_V -c_A \gamma^5 )u_s(p)\right]~.
\end{align}
where $c_V$ and $c_A$ are vector and axial-vector couplings. We now substitute this expression in Eq. \eqref{boltzmann1} and using this fact that for neutrinos
\begin{align}
\sum_{r}\bar{u}_r(q)\gamma^\mu(1-\gamma^5)u_r(q)=2q^\mu~,
\end{align}
we find the time evolution of polarization matrix as
\begin{align}\label{eq:30}
 \frac{d\rho^{(N)}_{ij}(\mathbf{k},t)}{dt}=i\frac{G_F}{4\sqrt{2}}\int d\mathbf{q} \,n_{\nu}(\mathbf{x,q})   \left(\delta_{is} \rho^{(N)}_{s'j}(\k,t)- \delta_{ js'} \rho^{(N)}_{is}(\k,t) \right)q^\mu\left[\bar {u}_{s'}(p')\gamma_{\mu} ( c_V -c_A \gamma^5 )u_s(p)\right]~.
\end{align}
To proceed, we define the average neutrino momentum as \cite{Mohammadi:2013ksa}
\beq
 \bar{\mathbf{q}}\,n_\nu(\mathbf{x})=\int \frac{d^3 q}{(2\pi)^3}\, \mathbf{q} \,n_\nu(\mathbf{x,q})~,
\eeq
where
\beq
n_\nu(\mathbf{x})=\int \frac{d^3 q}{(2\pi)^3} \,n_\nu(\mathbf{x,q})~.
\eeq
Therefore, the time evolution of fermion polarizations is simplified as
\begin{align}\label{zeta1}
\dot{\zeta}^{(N)}_{x}=-\frac{G_F}{\sqrt{2}}  |c_A| \frac{n_\nu(\mathbf{x})}{E_\nu}\left[ \bar{q}_3 \zeta^{(N)}_y+ \bar{q}_2 \zeta^{(N)}_z\right]~,\\ \label{zeta2}
\dot{\zeta}^{(N)}_{y}=\frac{G_F}{\sqrt{2}}  |c_A| \frac{n_\nu(\mathbf{x})}{E_\nu} \left[ \bar{q}_3 \zeta^{(N)}_x - \bar{q}_1 \zeta^{(N)}_z\right]~,\\ \label{zeta3}
\dot{\zeta}^{(N)}_{z}=\frac{G_F}{\sqrt{2}}  |c_A| \frac{n_\nu(\mathbf{x})}{E_\nu}\left[ \bar{q}_1 \zeta^{(N)}_y+ \bar{q}_2 \zeta^{(N)}_x\right]~.
\end{align}
In vector notation, one can write \cite{Schweiger:2001aq}
\beq \label{vectorB}
\frac{d}{dt}\bm{\zeta}^{(N)}= -\mu_N\, \bm{\zeta}^{(N)}\times \mathbf{ G}_{\textrm{eff}}~,
\eeq
where
\beq\label{MagneticField}
\mathbf{ G}_{\textrm{eff}}=|c_A|\frac{G_F}{\sqrt{2}}\frac{n_\nu (\mathbf{x})}{E_\nu}\frac{1}{\mu_N}(\bar{q}_1,-\bar{q}_2,\bar{q}_3)~.
\eeq
Physically in \eqref{vectorB}, the Stokes parameters of Nuclei effectively experience a static effective gradient field $\mathbf{ G}_{\textrm{eff}}$.
Although there is no external magnetic field to use NMR technic, the existence of an effective magnetic field can be verified by probing the time derivation of the Bloch vectors .

In the same manner, one can find the coherent torque exerted on the spinning fermions of a polarized medium due to forward scattering with neutrino  \cite{Moody:1984pt,Stodolsky:1974aq}.
 For $N$ fermions the magnetic moment per unit volume is $\mathbf{m}=N \bm{\mu}_N$ \cite{Pathria:2011ft,Weissbluth:1989aq}. For this system, the magnetic torque is given by
   \begin{align}\label{eq:41}
\bm{\tau}=N\bm{\mu}_N \times \mathbf{ G}_{\textrm{eff}}~.
\end{align}
where $\bm{\tau}$ as a time evolution of angular momentum is
\begin{align}\label{eq:42}
\bm{\tau}=-N g_N \frac{d}{dt}\braket{\mathbf{S}}=-N \frac{d}{dt}\bm{\zeta}^{(N)}~.
\end{align}
Combining (\ref{eq:42}) and the magnetic torque relation (\ref{eq:41}) and comparison with Eqs. \eqref{zeta1}-\eqref{zeta3} one can cheque the consistency of magnetic torque equation and (\ref{vectorB}).

\begingroup %------------------------------ BIBLIOGRAPHY
\makeatletter
\let\ps@plain\ps@empty
\makeatother
\bibliography{reference-fermions}
\endgroup

%===============================================================%
%************************* BIBLIOGRAPHY ************************%
%===============================================================%

%\begingroup %------------------------------ BIBLIOGRAPHY
%\makeatletter
%\let\ps@plain\ps@empty
%\makeatother
%\bibliography{reference}
%\endgroup
%\end{multicols}
\end{document}